\newcommand{\CLASSINPUTinnersidemargin}{18mm}
\newcommand{\CLASSINPUToutersidemargin}{12mm}
\newcommand{\CLASSINPUTtoptextmargin}{20mm}
\newcommand{\CLASSINPUTbottomtextmargin}{25mm}
\documentclass[10pt,conference,a4paper]{IEEEtran}

\usepackage[utf8]{inputenc}

\usepackage{times}

\usepackage{multirow}

\usepackage{graphicx}
\usepackage{subfigure}
\DeclareGraphicsExtensions{.png,.eps,.ps,.pdf}

\usepackage{url}

\usepackage[english]{babel}

\begin{document}

\title{Unveiling the I2P web structure: a connectivity analysis}

\author{\IEEEauthorblockN{Roberto Mag\'an-Carri\'on, Alberto Abell\'an-Galera, Gabriel Maciá-Fernández and Pedro Garc\'ia-Teodoro}
\IEEEauthorblockA{Network Engineering \& Security Group\\
Dpt. of Signal Theory, Telematics and Communications - CITIC \\
University of Granada - Spain\\
Email: rmagan@ugr.es, albertoabellan@correo.ugr.es, gmacia@ugr.es, pgteodor@ugr.es}
\and}

\maketitle

\begin{abstract}
Web is a primary and essential service to share information among users and organizations at present all over the world. Despite the current significance of such a kind of traffic on the Internet, the so-called Surface Web traffic has been estimated in just about 5\% of the total. The rest of the volume of this type of traffic corresponds to the portion of Web known as Deep Web. These contents are not accessible by search engines because they are authentication protected contents or pages that are only reachable through the well known as \textit{darknets}. To browse through darknets websites special authorization or specific software and configurations are needed. Despite TOR is the most used darknet nowadays, there are other alternatives such as I2P or Freenet, which offer different features for end users. In this work, we perform an analysis of the connectivity of websites in the I2P network (named \textit{eepsites}) aimed to discover if different patterns and relationships from those used in legacy web are followed in I2P, and also to get insights about its dimension and structure. For that, a novel tool is specifically developed by the authors and deployed on a distributed scenario. Main results conclude the decentralized nature of the I2P network, where there is a structural part of interconnected eepsites while other several nodes are isolated probably due to their intermittent presence in the network. 
\end{abstract}


\section{Introduction}

The quintessential Internet service, the World Wide Web (WWW), simply known as ``the Web'', has a hidden face which is ignored by a great number of users and organizations. The bulk of them browse the so-called \textit{Surface Web}, which corresponds to web resources able to be indexed by common search engines. However, the Surface Web is just the peak of an enormous WWW iceberg. The great part of the WWW contents are hosted on the so-called \textit{Deep Web}~\cite{bergman_white_2001}. These contents are not accessible by search engines mainly due to they are authentication protected contents or pages that are only reachable through the well known \textit{darknets}~\cite{al2019torank,kalpakis_osint_2016}. To browse through darknets websites a special authorization or specific software and configurations are needed.

Darknets have been widely publicized to users as technologies mainly intended to elude censorship restrictions imposed by totalitarian governments~\cite{hoang2018empirical}, and to preserve fundamental security rights like privacy or anonymity. Although this is true, the generalized use of darknets surpasses nowadays freedom demands to become a main tool for illegal actions and cibercriminality because of the impunity provided by them~\cite{Kumar2019}. 

Mainly motivated by that, darknets have received the attention of researchers in the last years. This way, some works in the literature have analyzed the content and services offered through this kind of technologies~\cite{biryukov2014content, al2017classifying, al2019torank}, as well as other relevant aspects like site popularity~\cite{biryukov2013trawling}, topology and dimensions~\cite{kadianakis2015extrapolating}, or classifying network traffic and darknet applications~\cite{montieri_dive_2020, montieri_anonymity_2020, cai_isanon_2019, yin_i2p_2019, shahbar_how_2018}. 

Two of the most popular darknets at present are {\it The Onion Router} (TOR; \url{https://www.torproject.org/}) and {\it The Invisible Internet Project} (I2P;\url{https://geti2p.net/en/}). This paper is focused on exploring and investigating the contents and structure of the websites in I2P, the so-called \textit{eepsites}. In particular, we are interested on how they are interconnected and which kind of relationships exist among them. Although some works have been proposed for TOR~\cite{avarikioti2018structure,sanchez2017onions} in this research line, no efforts can be found regarding I2P. In fact, we only found the work~\cite{gao2017large} where the authors attempt to discover and analyze eepsites claiming they discovered the 80\% of total eepsites in I2P. However, no eepsites' relationships and connectivity analysis are there provided.

To analyze the structure and connectivity of websites in I2P, a crawling and scrapping tool has been specifically developed here: \texttt{c4i2p} (Crawling for I2P), which is able to extract some useful information from eepsites to characterize them and how they are inter-connected. For that, a 3 months long experiment has been carried out in a distributed environment composed by different nodes located at different places. As we will discuss below in the document, around only $\sim 1.5\%$ of the total observed services correspond to eepsites. Besides, they are generally small and simple, composed of few pages with not so many text and hypertext. Moreover, about $\sim 66\%$ of the eepsites are isolated from the rest, thus comprising a hidden part of the overall set of eepsites. That means that I2P darknet can be seen as an heterogeneous and decentralized network, where also some popular sites exist. These popular eepsites are characterized by their persistence in the I2P network over the time, thus making a kind of eepsite network backbone. The rest of them seem to be appearing and disappearing randomly in/from the network as it can be seen through the present work.

The rest of the document is organized as follows. A main background on analysis of darknet related technologies is provided in Section~\ref{sec:estado_arte}. After that, Section~\ref{sec:fundamentos} introduces the main fundamentals of I2P and presents the c4i2p tool aimed at discovering the structure of eepsites. The experimental setup using the proposed tool is detailed in Section~\ref{sec:entorno_experimental}, while the results extracted from the experimentation are afterwards discussed in Section~\ref{sec:resultados}. Finally, the main conclusions and future work are presented in Section~\ref{sec:conclusiones}.

\section{Background on darknets}\label{sec:estado_arte}

The Surface Web or simply the Web, has been widely studied by the research community with different and heterogeneous aims. A lot of works have been focused on different related topics, such as performance optimization for search engines~\cite{spertus1997parasite,carriere1997webquery,lawrence1999accessibility,brin1998anatomy}, analysis of connectivity, dimension and structure of the sites~\cite{bharat1998improved,kleinberg1999web,broder2000graph,meusel2014graph} and classification of the sites and their contents~\cite{baeza2001relating,henzinger2001hyperlink, ding2004link}. 

Regarding the Deep Web, many works have focused their attention on darknets, specially on TOR. For instance, a generic crawling framework is proposed in~\cite{iliou2016hybrid} to discover resources with different contents hosted both on the Surface Web or the Deep Web. In particular, the authors carry out an experiment intended to search websites with content about homemade explosives. Another study focused on measuring performance parameters for TOR can be found in \cite{sochor2013fuzzy}. There, response time, throughput and latency are measured, among others parameters. The authors conclude that browsing TOR is slower than doing it through the Surface Web. In~\cite{biryukov2013trawling}, an approach to enumerate all hidden services in TOR is introduced. The authors take advantage of defects in both the design and implementation of TOR hidden services that could allow an attacker to deanonymize them. The analysis of popularity of hidden services in TOR is also carried out in \cite{biryukov2014content} and \cite{kadianakis2015extrapolating}. Both works are aimed at analyzing how much traffic is due to the use of hidden services and how many unique \texttt{*.onion} addresses exist in the network, in order to estimate the number of TOR hidden services. 

On the other hand, a  system for crawling the Deep Torrent, that is the torrents available in BitTorrent that cannot be found through public sites or regular search engines, is presented in~ \cite{rodriguez2017understanding}. Authors estimate what percentage of resources shared in the BitTorrent network are hidden or part of the Deep Web. In~\cite{avarikioti2018structure}, a very complete survey about the topology and content of TOR is presented. Authors make use of a specific crawler that recursively explores the links found. Using it, a total of 34,000 hidden services are found, 10,000 of them corresponding to online services. The work concludes that most of hidden services are well connected through central sites such as wikis and forums. That is, the structure of this network is somewhat centralized. Regarding contents, half of the sites involve lawful activities, while those with illegal hidden services are mainly related with fraud, sale of counterfeit products and drug markets. 

In~\cite{al2017detecting}, the authors make use of graphs to analyze the most popular emerging products in TOR. For that, authors make use of text information extracted from the domains corresponding to markets to create a Product Correlation Graph (PCG). In a similar line, a link inter-connectivity study is carried out in~\cite{sanchez2017onions} about the structure and privacy of TOR hidden services. The authors analyze more than 1.5 million URLs hosted in 7,257 TOR domains. For each page, links, resources and redirection graphics, as well as the language used, are analyzed. Very relevant conclusions are extracted, such as the fact that domains in TOR are highly interconnected and that there are many links within TOR pointing to pages hosted on the Surface Web (the reverse case is also quite common). 

In 2018, the creation of a database called DUTA was presented in \cite{al2017classifying}. The database contains a list of a multitude of TOR hidden services classified by content. It is noteworthy to notice that more than 250,000 addresses of hidden services were found, but only 7,000 of them were accessible while the rest were down or unavailable. More recently and in the same line of the previous work, a new algorithm called TORank is proposed in~\cite{al2019torank} to classify hidden services in TOR. Authors thoroughly analyze site contents, then creating a dataset (DUTA-10K) which updates DUTA. In addition, ToRank is assessed and quantitatively compared with some of the most popular classification algorithms, such as PageRank, HITS and Katz. Among the most interesting conclusions we can mention is the fact that only 20\% of the available hidden services refer to suspicious activities, and 48\% of them are associated with normal activities. They also discovered that, in general, domains related to suspicious activities have multiple clones under different locations (URLs), which is susceptible to be used as an additional feature to identify them. 

Some other works analyze the main weaknesses of TOR and I2P regarding anonymity that might compromise user identities and communication links. This is the case of ~\cite{erdin2015find}, where the authors conclude that some of the attacks require considerable resources to be effective and that, therefore, it is very unlikely that they will succeed against this kind of networks. As a consequence, both darknets are considered highly safe. A theoretical comparison between TOR and I2P is carried out in~\cite{ali2016tor} from a number of perspectives: visibility by the community, scalability, memory usage, latency, bandwidth, documentation, vulnerability to DoS attacks, number of exit nodes, etc. Authors in~ \cite{timpanaro2012bird} try to characterize the file-sharing environment within I2P, they evaluating how it affects the anonymity provided by the network. It is concluded that most of the activities within I2P are oriented to file sharing and anonymous web hosting. Moreover, it is also concluded that the nodes are geographically distributed.

Additionally, several works have been proposed based on the use of Machine Learning (ML) techniques and algorithms with the aim of analyzing and classifying darknet's network traffic in some sense. For instance, a ML-based approach to analyze traffic flows generated by I2P applications and users is introduced in~\cite{shahbar2017effects}. The work concludes that it is possible to create both user and application profiles, and that the accuracy in creating such profiles depends on the amount of shared bandwidth. More recently, the authors in~\cite{montieri_dive_2020} introduce a hierarchical machine learning based framework to classify the type of network, the type of traffic and the applications that generated it. For that, a labeled dataset (Anon17) comprising network traffic collected from I2P, TOR and JonDonym darknets is used. In comparison to their previous work~\cite{montieri_anonymity_2020} where they used a flat approach with several classic ML algorithm, they obtain an improved performance, specially in traffic classification. Moreover, the authors conclude that I2P applications are the hardest ones to classify in comparison to TOR. With the same aim, the authors in~\cite{cai_isanon_2019} proposed a three step XGBoost ML-based framework. As in the previous works, the Anon17 dataset is used to validate the solution. A relevant improvement, in terms of classification accuracy, is achieved in comparison to other works, e.g.,~\cite{shahbar2017effects}. Moreover, they also tested the suitability of the approach for classifying different but similar network traffic coming from Virtual Private Networks (VPN) or intentionally encrypted. Similar works, addressing the classification of I2P network data traffic are also found in the literature~\cite{yin_i2p_2019,shahbar_how_2018}.

The I2P network has also been analyzed from a forensic point of view in~\cite{wilson2016forensic}. This work is aimed to help a forensic analyst to find clues related to the activity of cybercriminals in I2P. More deeply, a study on the performance and safety of I2P is presented in~\cite{timpanaro2015evaluation}. The authors compare the design of the NetDB with the design of the popular KAD (Kademlia) algorithm. Among other conclusions, they stand out that I2P users are, in general, more stable than those of other non-anonymous P2P networks. Because of that, KAD design is considered less vulnerable to different attacks than the current NetDB design is.  Liu {\it et. al} introduce in~\cite{liu2014empirical} two passive and active methods to discover I2P routers. An experiment was carried out for two weeks, where around 25,640 routers were discovered every day. The authors claim that they almost cover 94.9\% of the I2P network compared to the data announced on the official website. A week-long experiment was also carried out to monitor the I2P network in~\cite{timpanaro2011monitoring}. Some interesting results were obtained in this study. Among others, the authors conclude that 37\% of the published leaseSets were offline after publication on NetDB. Around 30\% of the complete leaseSets set were not identified, which means that 30\% of the network was not running neither a web server nor an I2PSnark client. In addition, more than 50\% of the anonymous web services discovered were 70\% of the time available online. It should be noticed that the experiment had a relatively short duration and it was performed when the network was possibly not yet mature.


Another study is conducted in~\cite{liu2019i2p} to analyze I2P nodes. The authors collected 16,040 I2P nodes and analyzed some of their properties, such as distribution by country, bandwidth utilization and node FloodFill attributes. Finally, and perhaps one of the most complete academic studies on I2P to date, can be found in~\cite{hoang2018empirical}. In it, a comprehensive empirical study of the network is carried out where population, cancellation rate (abandonment of the network by users), type of router and geographic distribution of the I2P pairs are measured. At the time of the study, there were about 32,000 active I2P pairs in the network per day, and 14,000 of them were behind NAT or firewalls. Moreover, despite the decentralized nature of I2P, a censor actor could block more than 95\% of the peer IP addresses known by a stable I2P client by only controlling 10 routers in the network, what constitutes a serious deterioration of the connectivity of an I2P client. 

Through this review of the state of the art, we can conclude that the Surface Web has been explored and analyzed to a large extent from many points of view and approaches. We have focused the discussion on research examples mainly based on crawling, information gathering, search engines and graph-shaped connectivity. Although to a lesser extent than the Surface Web, TOR has also been studied in depth, from several points of view too, e.g., content, structure (best result of this analysis is the TorMetrics Project, \cite{TORMetrics}), etc. Nevertheless, I2P is still somehow unknown and further works about its operation and structure are desirable, specially for eepsites. In~\cite{gao2017large}, an attempt to discover and analyze eepsites is made. For that, authors propose the use of Floodfills, the active collection of \texttt{host.txt} files and the network crawling. A total of 1,861 online eepsites were discovered, this amount corresponding to 80\% of eepsites existing in this network as the authors claim. However, no eepsites' relationships and connectivity analysis is provided.  Because of that, the current work is mainly focused on the study of the I2P eepsites relationships and interconnections to unveil the I2P web structure. 

\section{I2P Darknet: Basics and Analysis}\label{sec:fundamentos}

In this section, we first present some fundamentals of I2P and afterwards describe the specific crawler developed to analyze that darknet.

\subsection{I2P fundamentals}\label{sec:fundamentos_i2p}

The I2P project~\cite{paginawebi2p} was started in 2003 by a group of hackers, developers and software architects as a useful set of tools against censorship, with anonymity and privacy in mind. 

I2P operation is similar to that in other anonymous networks. Thus, traffic is routed through various points in the network using chains of proxy servers. Each data packet is anonymously routed to the corresponding destination. Moreover, when users register their I2P instances (routers), these become \textit{nodes} of the network, thus contributing with their own network bandwidth to transport communications among other users. 

The use of I2P is accomplished by using applications specifically designed to interact with I2P in a transparent way. Two relevant examples of that are I2PSnark (client for BitTorrent), and I2PTunnel (services to route TCP/IP flows for common applications such as web browsers, IRC or SSH), among others.

One of the main elements that conform the I2P architecture are tunnels, since they allow sending (outbound tunnel) and receiving (inbound tunnel) messages between users/nodes in the network. A tunnel could be defined as the composition of a set of routers that are in charge of sending and receiving data packets to and from specific destinations or sources.


I2P makes use of a simple combination of symmetric and asymmetric encryption algorithms to provide data confidentiality and message integrity. It is commonly named as \textit{garlic} routing or encryption. More specifically, ElGamal/AES + SessionTags are used for the encryption with 2048-bit ElGamal, AES256, SHA256, and 32-byte nonces (single-use numbers).

The Network Database or NetDB is an important part of I2P as well. NetDB is a Distributed Hash Table (DHT) with a structure based on the Kademlia algorithm. It contains the information about I2P services so that nodes in the network can communicate with each other. NetDB is queried the first time a running instance of I2P wants to contact to another router in the network. NetDB is disseminated by means of the technique called Floodfill, which is based on the use of a series of special routers with the same name. Such special routers are in charge of distributing the database and synchronizing the corresponding information about the \textit{routerInfo} and the \textit{leaseSets} found on the network~\cite{bookdarknets}. Unlike TOR directory authorities, the Floodfill routers that compose the NetDB are not fixed or trusted routers. Indeed, any router on the network can become a Floodfill if it is configured in that way.

\subsection{C4i2p: A Crawling Tool for the I2P Darknet}\label{sec:herramienta_c4i2p}

Although some crawling tools have been developed to date with different aims \cite{ahmia}, none of them cover completely the main objectives this work is intended for: a connectivity analysis among eepsites. For this reason, a new and ready-to-use tool has been devised and developed: \texttt{c4i2p}, which stands for 'crawling for I2P'.  Beyond its usefulness to achieve the objectives of the current work, we are convinced about its potential use by the research community to be improved or extended in many ways. Because of that, c4i2p has been conceived as an open source project and thus it is publicly available for downloading at~\cite{repositorioc4i2p}.

Through the rest of this section, the main modules, processes and components that shape the developed tool are introduced and described.

\vspace*{0.3cm}
\subsubsection{Modules}

As mentioned before, c4i2p tries to be a specific tool that allows users to interact with the I2P network. Thanks to this tool, relationships among eepsites can be observed by gathering different kind of information. For that, several modules are involved in the operation of c4i2p, as shown in Fig.~\ref{fig:c4i2p_modules}:

\begin{figure*}[t]
	\centering
	\includegraphics[width=0.8\textwidth]{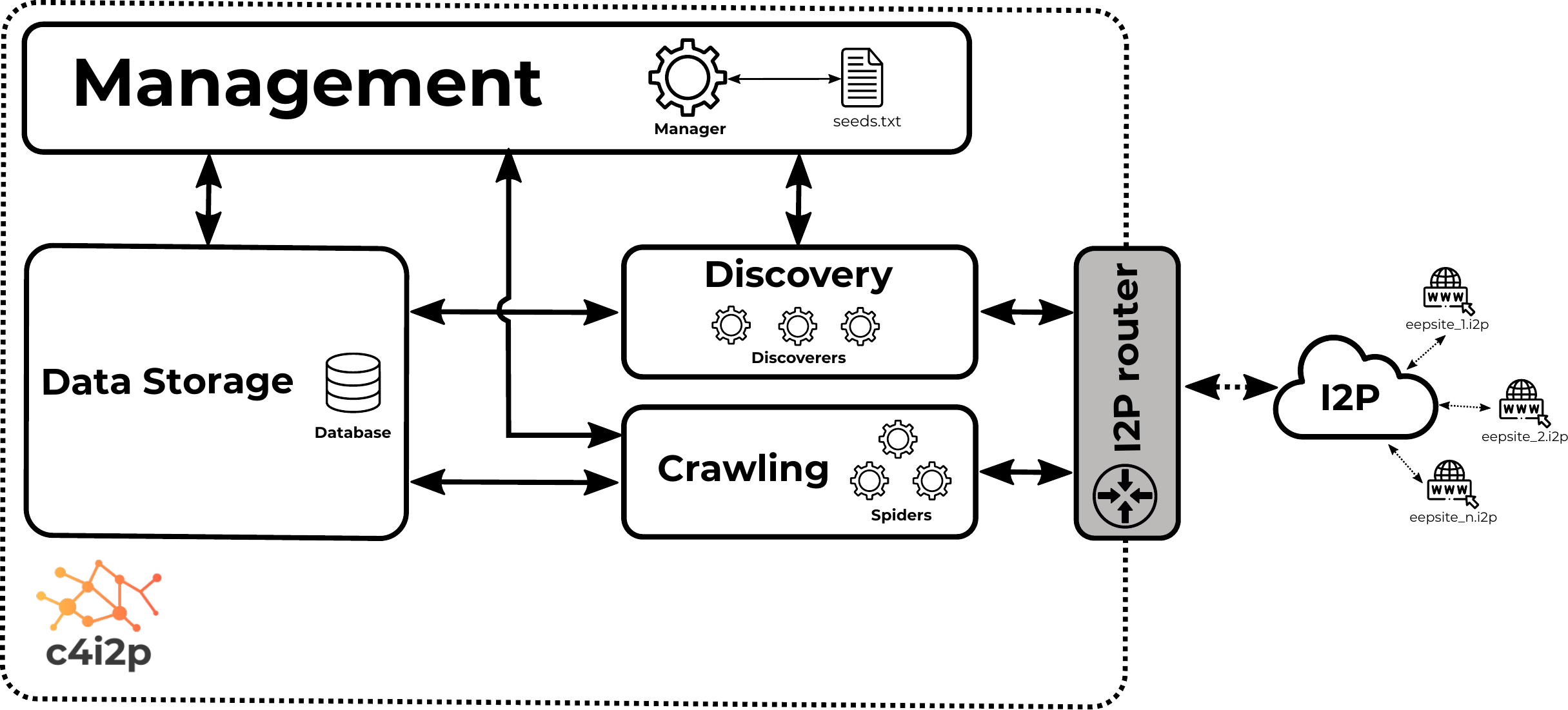}
	\caption{c4i2p modules and interactions.}
	\label{fig:c4i2p_modules}
\end{figure*}

\begin{itemize}
    \item \textbf{Management}. It is in charge of controlling the entire workflow of the tool. It governs the interaction between the rest of the procedures and elements composing the system. The management module is in turn managed by the end user.
    \item \textbf{Discovery}. It is responsible for checking the availability of the observed eepsites in a controlled manner over time.
    \item \textbf{Crawling}. It is responsible for carrying out the task of crawling and everything related with searching for and scrapping specific information from I2P eepsites.
    \item \textbf{Data Storage}. It is a module aimed to provide data stora\-ge services to others. It is responsible for the storage, persistence and management of all the information.
\end{itemize}

\vspace*{0.3cm}
\subsubsection{Processes, components and relationships}\label{sec:herramienta_c4i2p:elements}

The main components that intervene in the operation of c4i2p are as follows (see Fig.~\ref{fig:c4i2p_modules}):

\begin{itemize}
    \item \textbf{Sources}. They are eepsite addresses from I2P. They can be obtained from three different origins:
    \begin{itemize}
        \item SEED. It is obtained from the list of initial seeds (\texttt{seeds.txt}) from which the experiment starts. They were selected from a preliminary search for eepsites in I2P.
        \item FLOODFILL. They are discovered by the Floodfill I2P deployed routers.
        \item DISCOVERED. The ones discovered through the crawling process, i.e. from links obtained from each scrapped eepsite.
    \end{itemize}
    \item \textbf{Manager}. The main procedure of the Management module, the Manager is the brain and core of the entire tool. It is in charge of coordinating and orchestrating the rest of elements and to keep track of eepsite's status and lifecycle.  
    \item \textbf{Discoverers}. As the main part of the Discovery module, they implements the main functionality of the discovering procedure. Each one is responsible for checking if an eepsite is active (reachable and operational). How many of them and when they are launched is controlled by the Discovery module.
    \item \textbf{Spiders}. Each of them, separately, is in charge of carrying out the Crawling process of a certain \textit{eepsite}. An spider dives recursively into all the pages to extract the desired information. It is launched right after an eepsite is discovered.
    \item \textbf{Database}. The database stores in an ordered and structured way all the information extracted from eepsites, as well as the inherent data of the system itself.
    \item \textbf{I2P network}. The I2P network is where the eepsites are hosted, which are the entities from which we want to extract information.
    \item \textbf{I2P router}. Finally, the I2P router allows c4i2p to communicate with the I2P network.
    
\end{itemize}

\vspace*{0.3cm}
\subsubsection{Eepsites functional states}

During its lifecycle in c4i2p, an eepsite goes through several states. The states and transitions among them are depicted in Fig.~\ref{fig:diseno5}:

\begin{figure}[t]
	\centering
	\includegraphics[width=0.5\textwidth]{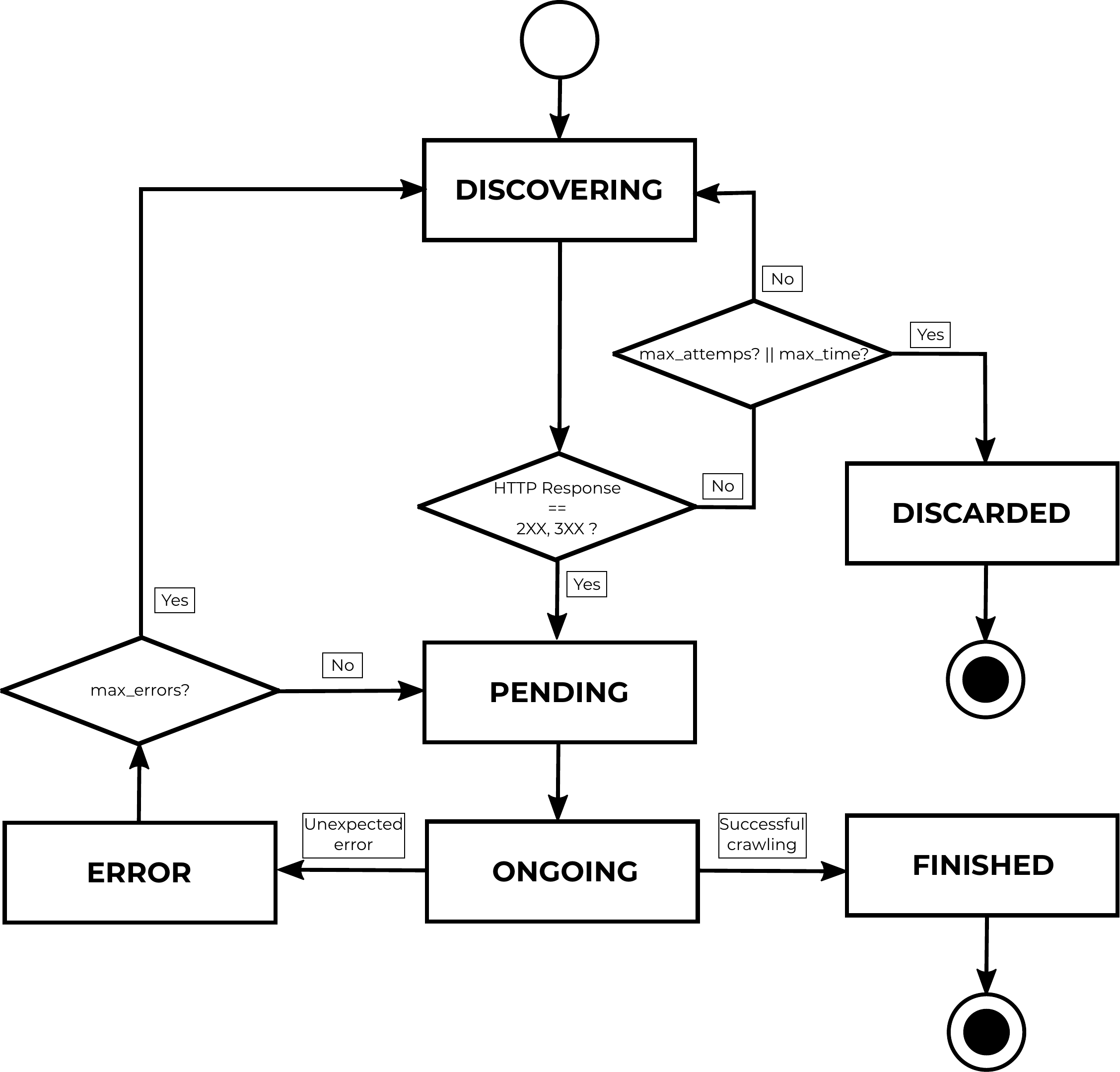}
	\caption{Eeepsite's lifecycle in c4i2p, where states (squared boxes) and available transitions (polygonal shapes) among them are depicted.}
	\label{fig:diseno5}
\end{figure}

\begin{itemize}
    \item \textbf{DISCOVERING}. When c4i2p takes a new eepsite from the list of sources until the eepsite is successfully contacted, the eepsite is being discovered. This is the initial state of every observed eepsite.
    \item \textbf{PENDING}. Once an eepsite is successfully contacted, i.e., it is reachable and available, it goes to the pending status where it is awaiting to be crawled. Indeed, the eepsite is queued into a FIFO based queue for that.
    \item \textbf{ONGOING}. The eepsite is being crawled.
    \item \textbf{ERROR}. An eepsite is in the ERROR state if there was an error while crawling it. In order to avoid not contemplated errors during the crawling process, the application allows a certain number crawling processes until the eepsite is proposed to be discovered again. 
    \item \textbf{FINISHED}. A node is labeled FINISHED if it was successfully crawled.
    \item \textbf{DISCARDED}. An eepsite reaches this state after a certain number of discovering attempts or when a predefined on-discovering time is exceeded. This state concludes that the eepsite is no longer available or it never existed.
\end{itemize}

\section{Experimental setup}\label{sec:entorno_experimental}

\begin{figure}[t]
	\centering
	\includegraphics[width=0.5\textwidth]{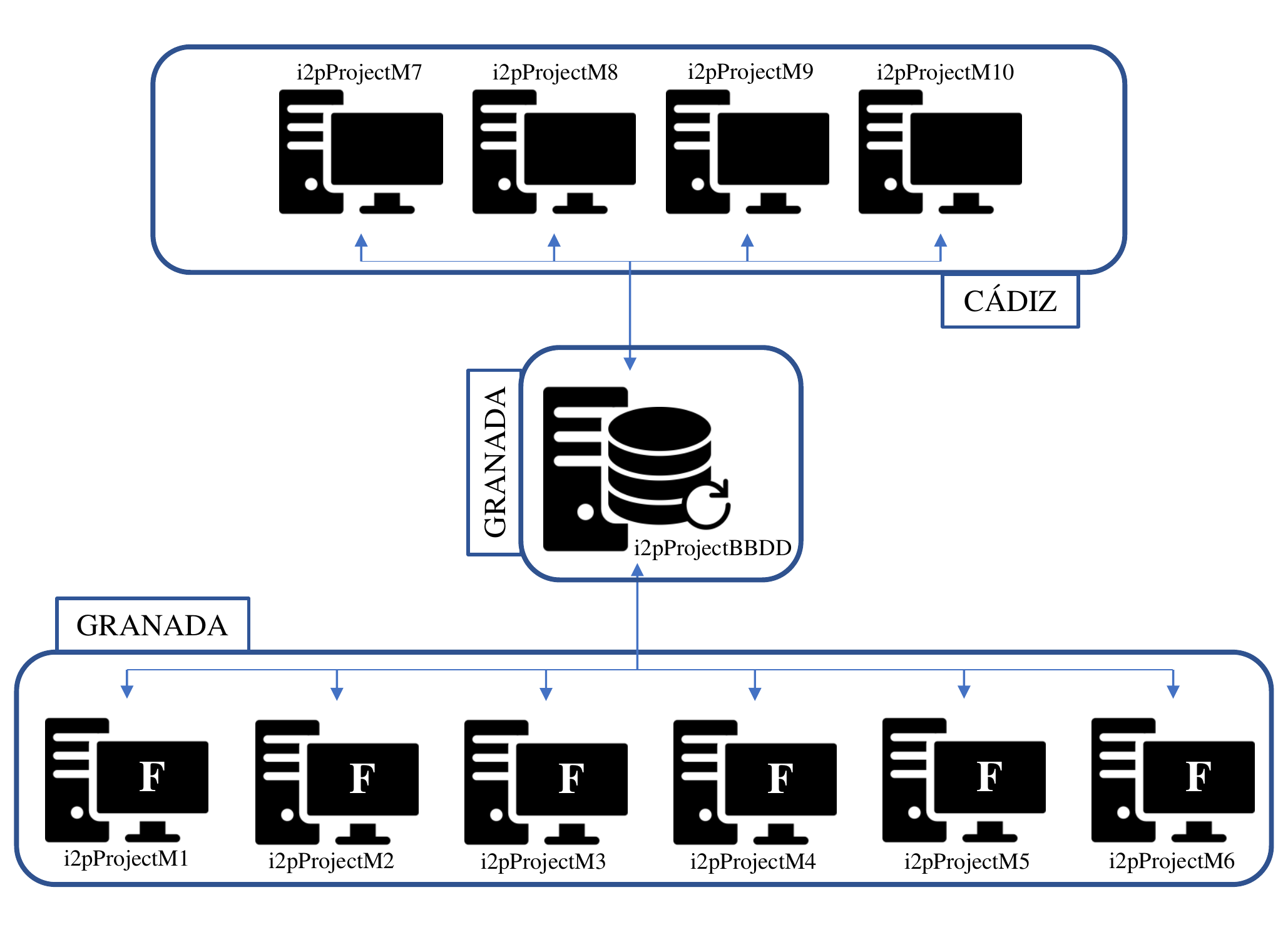}
	\caption{Experimental setup where 11 virtual machines are deployed and distributed in different locations: \texttt{i2pProjectM[1-6]} and \texttt{i2pProjectBBDD}, in Granada (Spain); and \texttt{i2pProjectM[7-10]}, in C\'adiz (Spain). All the machines (except that corresponding to BBDD) run an I2P router. The ones with IDs \texttt{i2pProjectM[1-6]} are configured as floodfill I2P routers.}
	\label{fig:entornoexperimental}
\end{figure}

In order to crawl the I2P network a distributed experimental environment is devised. It is composed of 11 virtual machines organized in two separated groups as shown in Fig.~\ref{fig:entornoexperimental}. They are distributed in different geographical places running in different computation clusters too. On the one hand, 7 of the VMs are deployed on the facilities of the University of Granada (Spain): 6 named as \texttt{i2pProjectM[1-6]} in the figure, and the one acting as the central database identified as \texttt{i2pProjectBBDD}. On the other hand, the remaining 4 VMs are deployed on the facilities of the University of Cádiz (Spain), which are identified in the figure as \texttt{i2pProjectM[7-10]}. All the virtual machines run Linux Ubuntu 18.04.1 LTS distribution with a total of 5 GBs RAM, 100 GBs HD and two 2.40 GHz vCPUs.

Every VM executes an I2P router instance for accessing the I2P darknet, except for the \texttt{i2pProjectBBDD} machine. Machines \texttt{i2pProjectM[1-6]} were set up as floodfill I2P routers, while the rest \texttt{i2pProjectM[7-10]} run as normal I2P routers. Floodfill routers, as mentioned in Section~\ref{sec:fundamentos_i2p}, allow to discover additional eepsites since they all are in charge of maintaining and managing the NetDB. Following recommendations from work \cite{Khan2018}, we set up shared router's communication bandwidth to 8 MBps, the number of maximum participants in tunnels being equal to 10K. In addition, it is necessary to activate floodfill routers by setting up \texttt{router.floodfillParticipant=true} in \texttt{router.config} configuration file.

All VMs also include a c4i2p instance. They altogether provide a distributed crawling procedure where each c4i2p instance is in charge of managing and discovering its own eepsites. Both, Discovery and Crawling modules were accordingly set up in each c4i2p instance. Among other configuration parameters, the number of simultaneous running spiders is limited to 10 to efficiently manage the VM computation capacities. Moreover, a maximum value is established for the discovery time and the number of attempts. If one of such limits is reached, the site in process of being discovered will be discarded. At the beginning, some initial I2P URLs (called {\it seeds}) are equally distributed among all c4i2p instances. In our experiment we initially count on 3,938 seeds, which means that $\sim$ 394 initials seeds are assigned to each instance. Table \ref{tab:c4i2p_config_details} shows the configuration parameters of c4i2p.

\begin{table*}[t]
    \centering
    \begin{tabular}{|c|c|l|}
    \hline
    \textbf{Parameter}                         & \textbf{Value} & \textbf{Description}                                 \\ \hline
    \texttt{MAX\_ONGOING\_SPIDERS}             & 10             & Number of simultaneous spiders               \\ \hline
    \texttt{MAX\_CRAWLING\_ATTEMPTS\_ON\_ERROR}   & 2              & Number of crawling attempts per eepsite                 \\ \hline
    \texttt{MAX\_DISCOVERY\_ATTEMPTS}           & 30d $\times$ 24h       & Number of discovery attempts per eepsite: one per hour during a month              \\ \hline
    \texttt{MAX\_DISCOVERY\_DURATION (m)}    & 30d $\times$ 24h $\times$ 60m       & Maximum time for which an eepsite is tried to be discovered \\ \hline
    \texttt{MAX\_DISCOVERY\_SINGLE\_THREADS} & 50             & Number of eepsites to be simultaneously discovered    \\ \hline
    \texttt{HTTP\_TIMEOUT (s)}                 & 30s             & HTTP request timeout from an I2P URL           \\ \hline
    \texttt{INITIAL\_SEEDS}                    & \texttt{seeds.txt} & Path to the file containing the set of eepsite URLs    \\ \hline
    \texttt{INITIAL\_SEEDS\_BATCH\_SIZE}        & $\sim$394            & Number of initial seeds assigned to a c4i2p instance \\ \hline
    \end{tabular}
    \caption{c4i2p configuration parameters.}
    \label{tab:c4i2p_config_details}
\end{table*}

\begin{figure*}[t]
	\centering
	\includegraphics[width=1\textwidth]{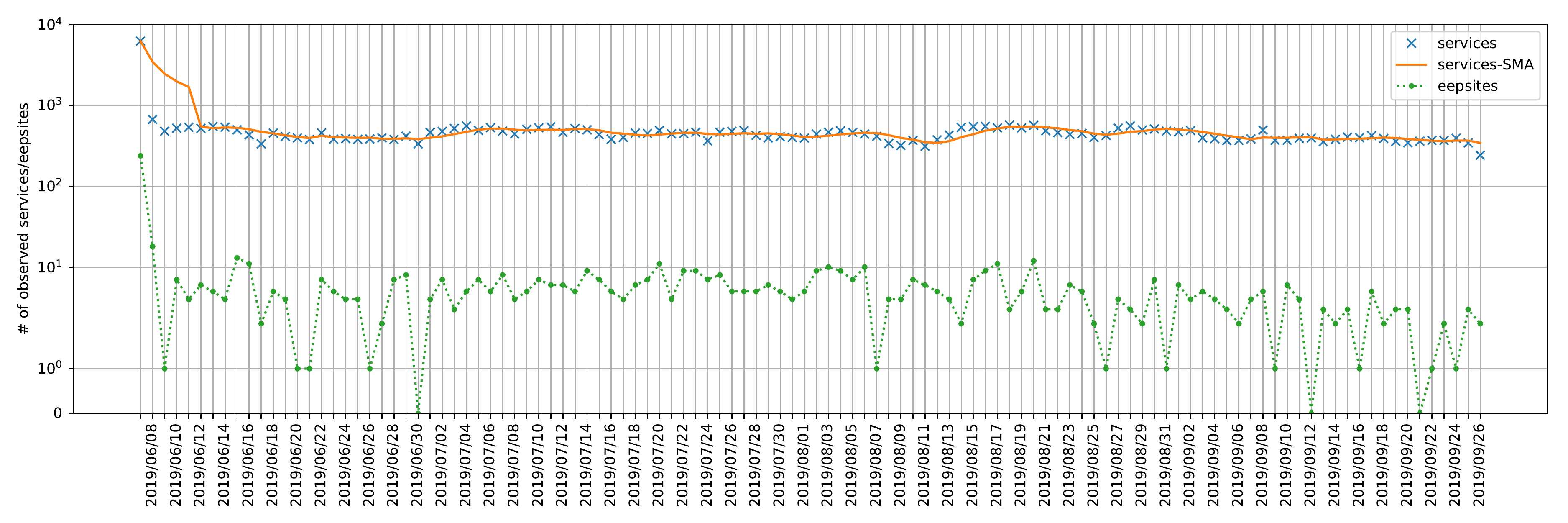}
	\caption{Daily number of observed I2P services (blue crosses), SMA (Simple Moving Average) (orange line) and eepsites (green dots). Notice the high number of services/eepsites at the beginning of the experiment where the system added more than 6,000 I2P services from initial seeds and floodfill sources.}
	\label{fig:total_services_vs_eepsites}
\end{figure*}

\begin{figure*}[t]
	\centering
	\includegraphics[width=1\textwidth]{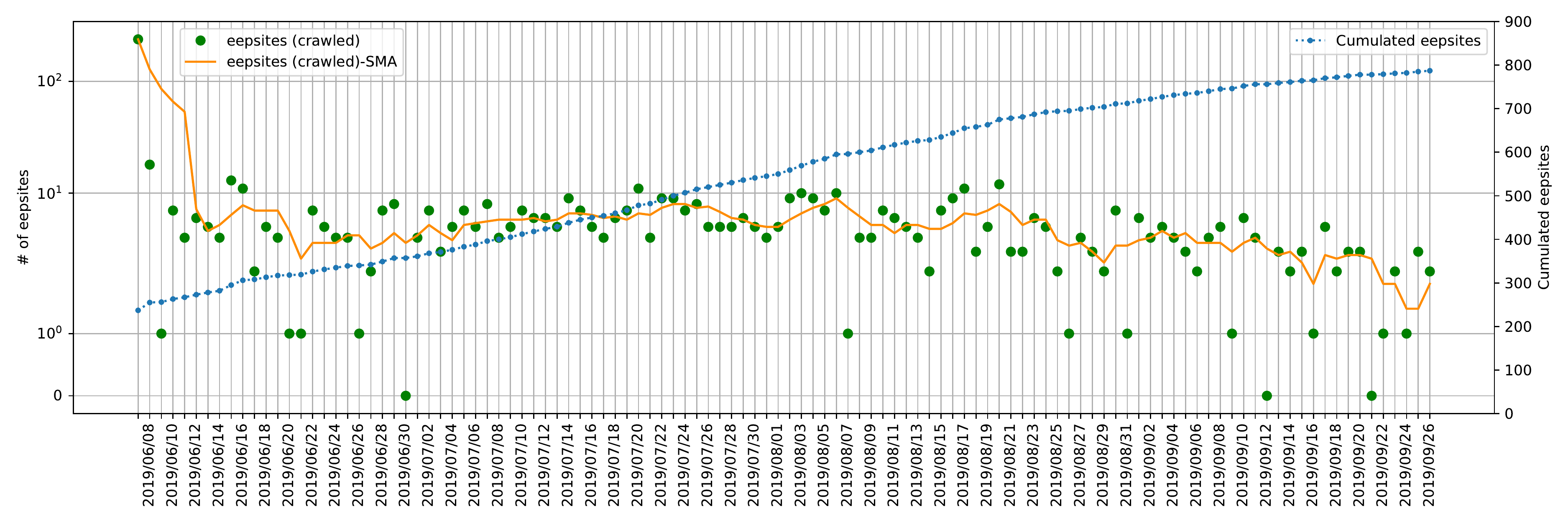}
	\caption{Detailed view of the daily rate of eepsites found and crawled (green dots), SMA (Simple Moving Average) (orange line) and cumulated sum (blue crosses).  At the beginning of the experiment, the system crawled more than 200 eepsites from seed, floodfill and discovered sources.}
	\label{fig:total_eepsites}
\end{figure*}

Finally, the \texttt{i2pProjectBBDD} machine runs a MySQL DataBase Management System (DBMS) to provide data persistence and storage for the rest of the crawling instances. The DBMS engine has to be configured to support a high number of concurrent connections. As it can be seen in Table \ref{tab:c4i2p_config_details}, every c4i2p instance simultaneously runs 50 discovery threads (the discoverers), the DBMS engine being configured to support a minimum of 10 (VMs) $\times$ 50 (threads) = 500 concurrent connections.

\section{Results on I2P Eppsites and Discussion}\label{sec:resultados}

The experiment took a total of 111 days, it starting on June 7th 2019 and finishing on September 26th 2019. During this period we managed a total of 54,974 I2P URLs, most of them obtained from leasesets of floodfill routers. They correspond to services that can be web related (eepsites) or not. In fact, only 787 of them were eepsites and, thus, they were successfully crawled. 

Figures~\ref{fig:total_services_vs_eepsites} and~\ref{fig:total_eepsites} show the number of daily observed services during the experiment. As expected, the number of eepsites found (green dots in the figures) is much lower than the number of total observed services in the darknet (blue crosses in the figures). In Figure~\ref{fig:total_eepsites}, a detailed view of the total number of eepsites is shown together with their cumulated sum (blue dots in the figure). Additionally, a SMA (Simple Moving Average) for 5 days (continuous orange line in both figures) is computed in order to see trends in the discovering and crawling process.

As it can be seen in the previous figures, new I2P services are found every day. Indeed, a daily average of $\sim 500$ services is computed at the end of the experiment. Similarly, eepsites are crawled almost every day, that resulting in an average of $\sim 5$ which motivates the linear increment in the number of found eepsites from the beginning of the experiment shown in Figure~\ref{fig:total_eepsites} (blue dots). The high number of eepsites observed in the first day is motivated by the initial system contribution to the total number of services. These eepsites come from seed, floodfill and discovered sources. Moreover, no significant trends can apparently be seen for both total services and eepsites, according to the SMA values.

\begin{figure}[t]
	\centering
	\includegraphics[width=0.5\textwidth]{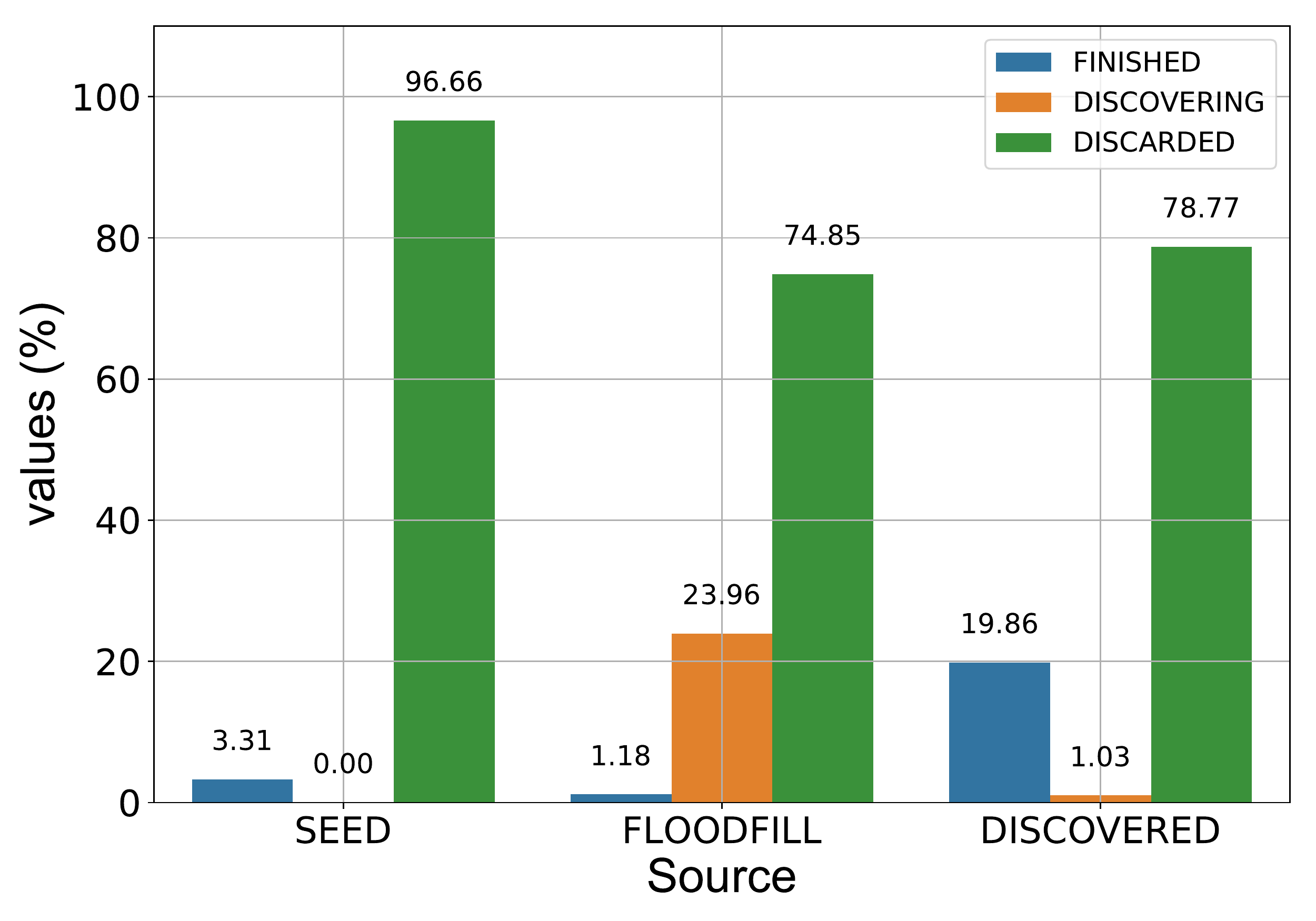}
	\caption{Source vs. status results: Percentage of services from a specific source in a specific status.}
	\label{fig:sourcestatus}
\end{figure}

Some interesting conclusions can be also obtained from the analysis of the status and sources of the observed services. Figure~\ref{fig:sourcestatus} shows the rate of observed services in a specific status, ordered by their source. These results are numerically shown in Table~\ref{tab:sourcesstatus} too. From them, it can be seen that the number of discarded eepsites is very high in comparison with the ones finally crawled (finished). Moreover, the number of finished eepsites is only $\sim 1.5\%~(787)$ of the total observed services in the darknet. Notice that most of the services were obtained from floodfill routers (50,784), the number of discovered ones being really low (292). This last value is clearly motivated by the low number of eepsites finally crawled. However, in percentage, the number of finished eepsites obtained by crawling other eepsites (discovered) is higher than those obtained from floodfill routers. Additionally, as expected, the higher number of discovered eepsites comes from floodfill sources. Finally, $96.66 \%$ of services from seed related sources have been discarded. The initial list of seeds was chosen manually from a previous work but, as the results shown, it can be obsolete with most of the eepsites no longer available currently. As a consequence, we can conclude that the number of eepsites comprising the I2P network changes over time, some of them disappearing from the network while some others are created.

\begin{figure}[t]
	\centering
	\includegraphics[width=0.5\textwidth]{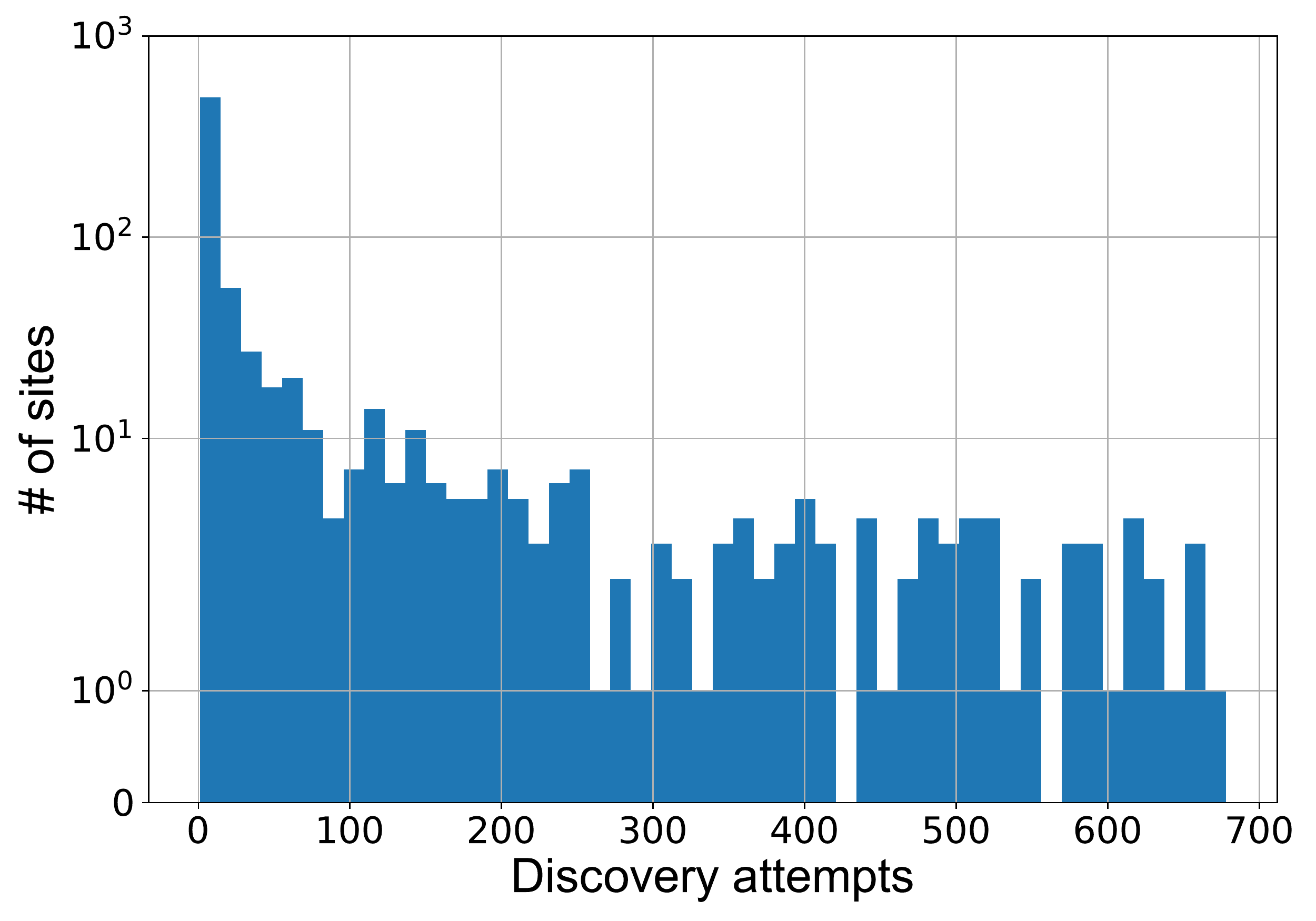}
	\caption{Distribution of discovery attempts for finished eepsites.}
	\label{fig:discoveringtries}
\end{figure}

\begin{table*}[!t]
\centering
\begin{tabular}{|c|c|c|c|c|c|c|}
\hline
\textbf{Source} &
  \textbf{\# services by source} &
  \textbf{Status} &
  \textbf{\# by status} &
  \textbf{\% by status} &
  \multicolumn{1}{l|}{\textbf{\# total services}} &
  \textbf{\% of total services} \\ \hline
\multirow{3}{*}{SEED}       & \multirow{3}{*}{3,938}  & FINISHED    & 129   & 3.31   & \multirow{9}{*}{54,974} & 0.25  \\ \cline{3-5} \cline{7-7} 
                            &                        & DISCOVERING & 0     & 0      &                        & 0     \\ \cline{3-5} \cline{7-7} 
                            &                        & DISCARDED   & 3,768  & 96.66  &                        & 6.85  \\ \cline{1-5} \cline{7-7} 
\multirow{3}{*}{FLOODFILL}  & \multirow{3}{*}{50,784} & FINISHED    & 600   & 1.18   &                        & 1.09  \\ \cline{3-5} \cline{7-7} 
                            &                        & DISCOVERING & 12,167 & 23.96  &                        & 22.13 \\ \cline{3-5} \cline{7-7} 
                            &                        & DISCARDED   & 38,012 & 74.85  &                        & 69.14 \\ \cline{1-5} \cline{7-7} 
\multirow{3}{*}{DISCOVERED} & \multirow{3}{*}{292}   & FINISHED    & 58    & 19.86  &                        & 0.15   \\ \cline{3-5} \cline{7-7} 
                            &                        & DISCOVERING & 3     & 1.03   &                        & 0.005 \\ \cline{3-5} \cline{7-7} 
                            &                        & DISCARDED   & 230   & 78.77  &                        & 0.42  \\ \hline
\end{tabular}
\caption{Source vs. status results: Aggregated results obtained from the experiment depending on the source and the eepsites' status.}
\label{tab:sourcesstatus}
\end{table*}

It is also interesting to see the distribution of the number of discovering attempts consumed for finished eepsites. This is shown in Figure~\ref{fig:discoveringtries}. A deep inspection of the results concludes that only $\sim 20 \%$ of eepsites needed 1 attempt to be finally crawled. From that we can conclude the poor availability of eepsites in the I2P network. Another interesting result is the fact that some other nodes required an exceptional number of attempts to be finally crawled. Indeed,  $\sim 13 \%$ of eepsites needed more than or equal to 200 attempts to be crawled. The main hypothesis behind this result is that these sites are continuously appearing and disappearing from the network and they are difficult to be reached with just few discovering attempts.

In summary, new I2P services are continuously appearing in the darknet though most of them are not eepsites and have been obtained from floodfill routers. Moreover, most of active eepsites in the past, those included as seed sources in this experiment, are not currently available. This could mean that, in general, eepsites have a short life. In fact, only $3.31 \%$ of the eepsites from seed sources have been finally crawled according to Table~\ref{tab:sourcesstatus}. As it will be seen in the next section, such eepsites correspond to either \textit{source} or \textit{sink} eepsites (nodes) in the darknet. Moreover, even having a low number of finished eepsites from discovered sources, it is expected to find a higher number of eepsites obtained from crawling eepsites (discovered) than from other kind of sources due to their interconnections and relationships. Finally, active and persistent eepsites are easy to be discovered.

\subsection{Size analysis}

After the various results and behaviors previously described, we now perform an analysis about the size of I2P eepsites. For that, information about their number of pages is extracted during the crawling procedure. Additionally, the number of words, letters, images and scripts in the home page, as well as the predominant eepsite's language, are also estimated.

\begin{figure}[t]
	\centering
	\includegraphics[width=0.5\textwidth]{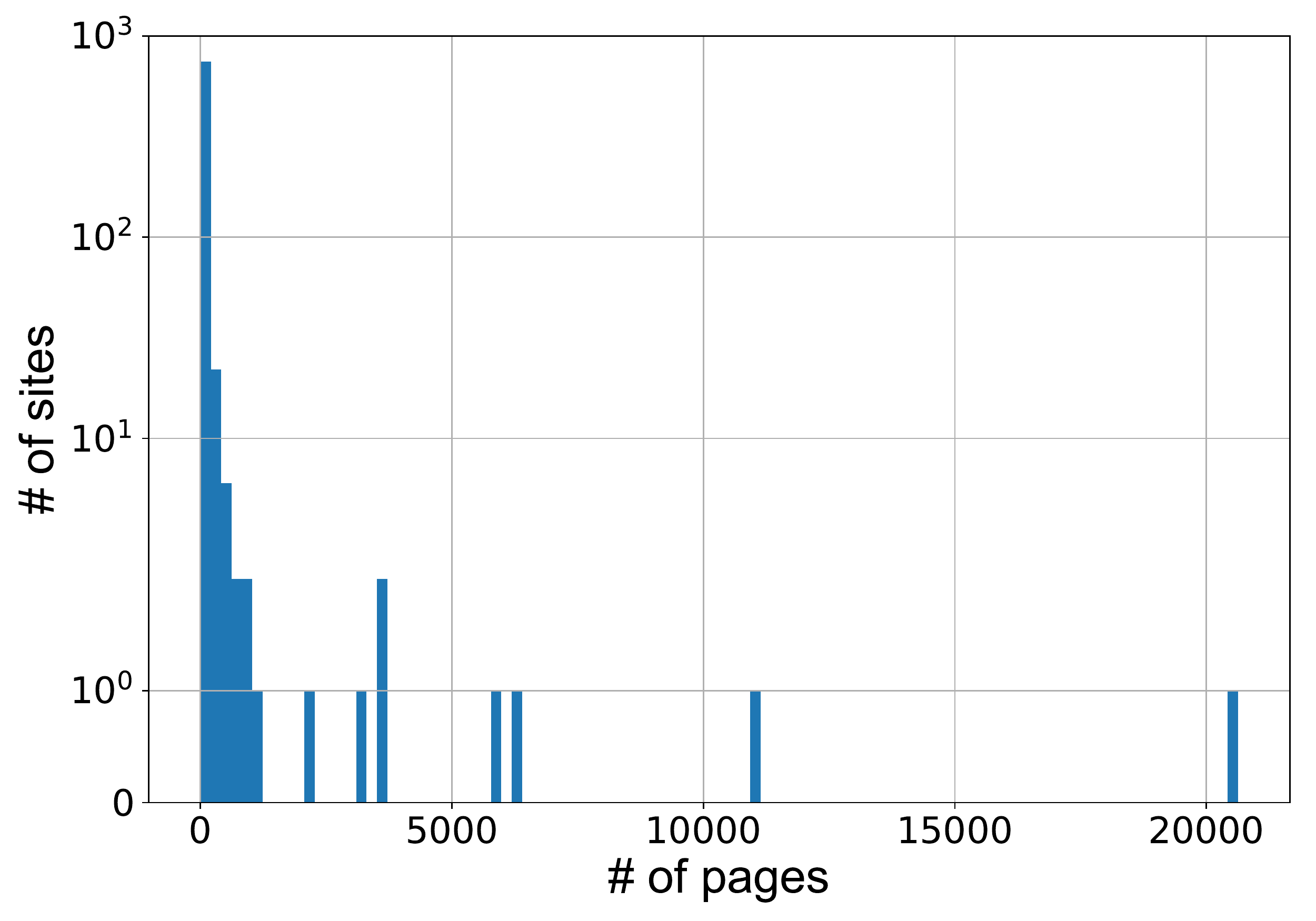}
	\caption{Eepsites size: Distribution of the number of pages of found eepsites.}
	\label{fig:pages_finished}
\end{figure}

According to Figure~\ref{fig:pages_finished}, the I2P darknet is mostly composed of small eepsites. In fact, $\sim 80 \%$ of the total number of eepsites have less than or equal to 30 pages. However, there are some exceptions. For instance, extraordinary large eepsites with 5, 6 or even 11 thousand pages exist, the largest one having 20,630 pages. Table~\ref{tab:largest_pages} shows sites with more than 1,000 pages.

\begin{table*}[t]
    \centering
    \begin{tabular}{|c|c|}
    \hline
        \textbf{\# of pages} &                                          \textbf{Eepsite} \\
    \hline
      20630 &  \texttt{ux6prousphswf56bym7yo7kst4ybh45y2z2wrnw7dujmrz56hq4q.b32.i2p} \\ \hline
      11059 &  \texttt{qiii4iqrj3fwv4ucaj2oykcvsob75jviycv3ghw7dhzxg2kq53q.b32.i2p} \\ \hline
      6336 &  \texttt{qa4boq364ndjdgow4kadycr5vvch7hofzblcqangh3nobzvyew7a.b32.i2p} \\ \hline
      5913 &  \texttt{whba2ljn2sjvke45yjkyudzmelwkjcop3m7r6kubohngq3pb6cqa.b32.i2p} \\ \hline
      3614 &  \texttt{a2lnpfsrhy5d3yky6xsut6gj6j76vn3lsy7kvabvedtu2d37s65q.b32.i2p} \\ \hline
      3516 &  \texttt{gwqdodo2stgwgwusekxpkh3hbtph5jjc3kovmov2e2fbfdxg3woq.b32.i2p} \\ \hline
      3118 &  \texttt{u6pciacxnpbsq7nwc3tgutywochfd6aysgayijr7jxzoysgxklvq.b32.i2p} \\ \hline
      2186 &  \texttt{i2pforum.i2p} \\ \hline
      1194 &  \texttt{25cb5kixhxm6i6c6wequrhi65mez4duc4l5qk6ictbik3tnxlu6a.b32.i2p} \\ 
    \hline
    \end{tabular}
    \caption{I2P eepsites with more than 1,000 pages.}
    \label{tab:largest_pages}
\end{table*}

\begin{figure*}[!hbt]
	\begin{centering}
		\subfigure[]{
			\includegraphics[scale=0.325]{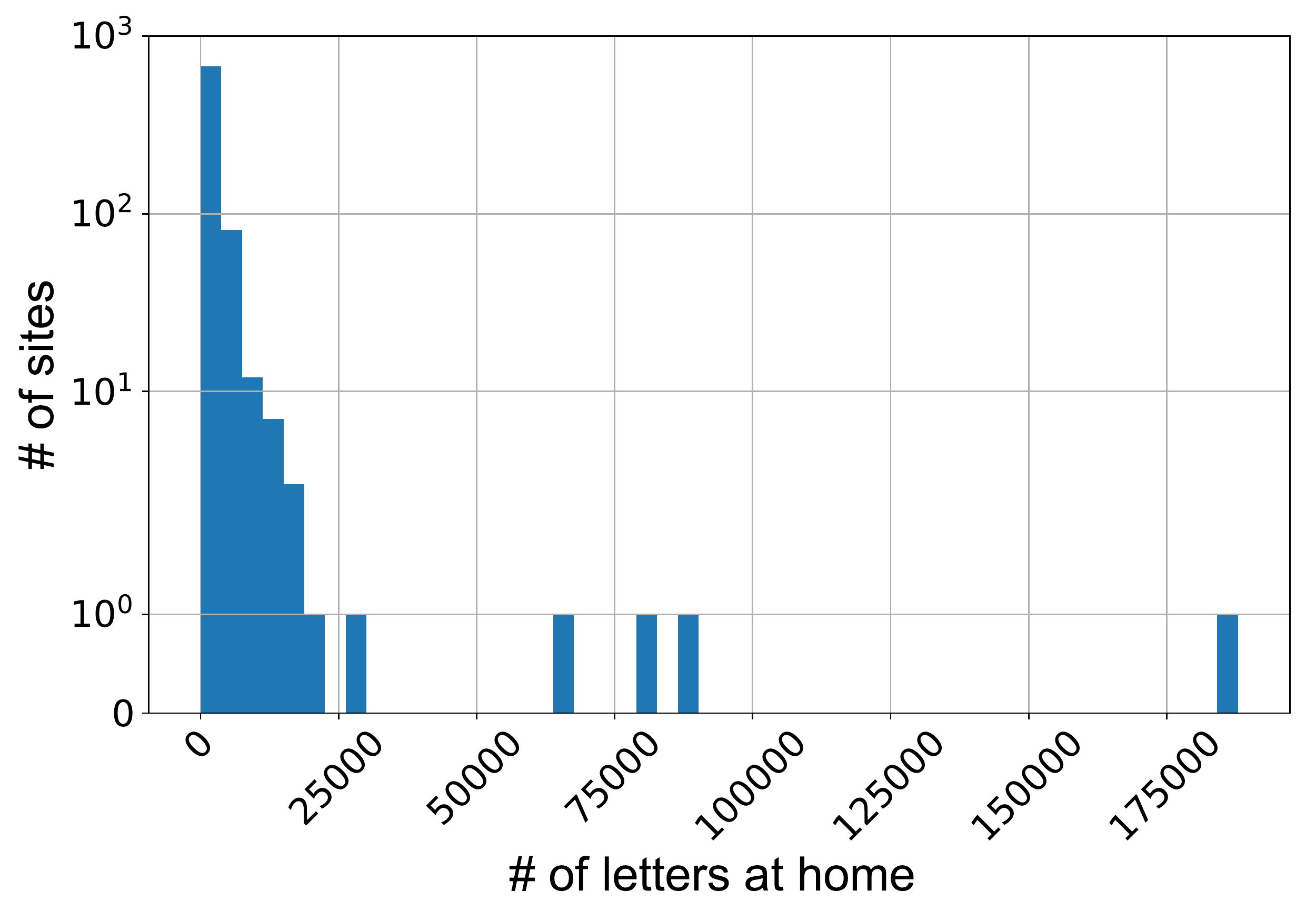}
			\label{fig:content:letters}
		}
		\subfigure[]{
			\includegraphics[scale=0.325]{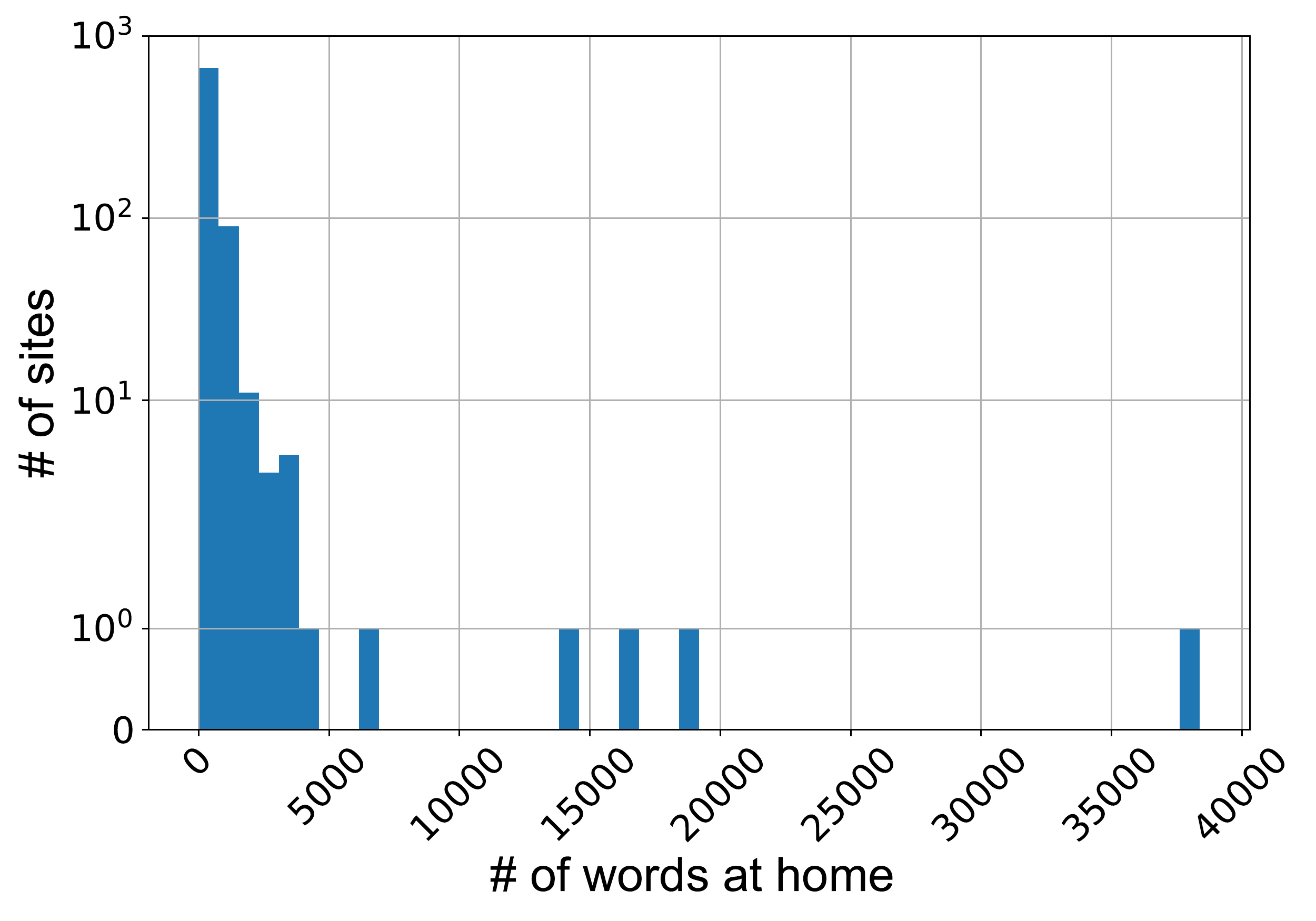}
			\label{fig:content:words}
		}
		\subfigure[]{
			\includegraphics[scale=0.325]{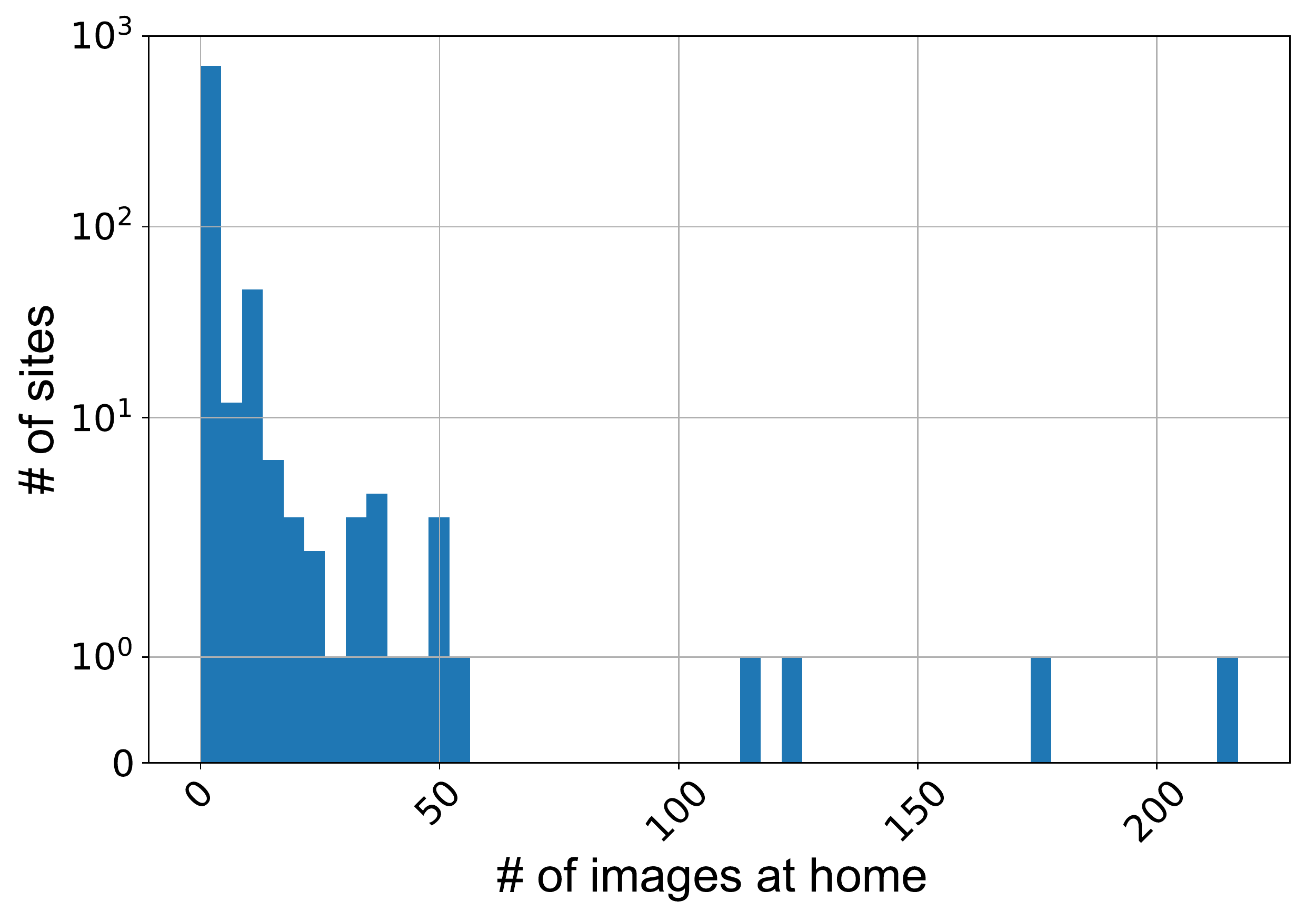}
			\label{fig:content:images}
		}
		\subfigure[]{
			\includegraphics[scale=0.325]{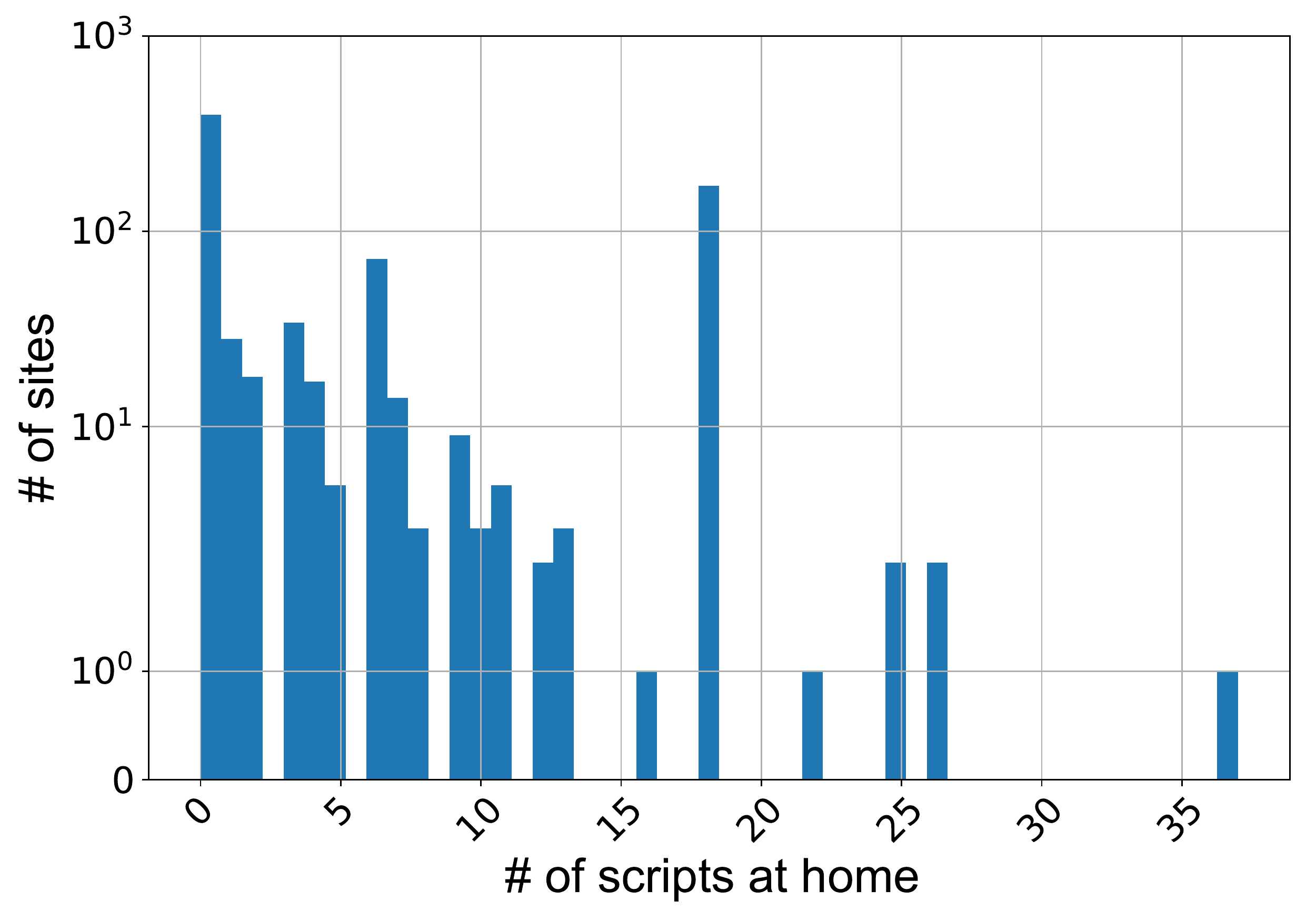}
			\label{fig:content:scripts}
		}
		\caption{Eepsites content analysis in terms of number of \subref{fig:content:letters} letters, \subref{fig:content:words} words, \subref{fig:content:images} images and \subref{fig:content:scripts} scripts.}
		\label{fig:content}
	\end{centering}
\end{figure*}

Regarding the size of eepsites related pages, Figure~\ref{fig:content} shows that almost all eepsites have a home page with a few number of letters, words, images and scripts. 



Finally, also an analysis about the predominant language used in eepsites is performed. To do that, we considered a widely accepted language detection engine: Google Translator. The results obtained are shown in Table~\ref{table:idiomas}, from which we can conclude that English is the most used language. However, we cannot be very confident about these results due to the small number of crawled eepsites.

\begin{table}[t!]
\centering
\caption{Language of eepsites according to Google Translator engine.}
\begin{tabular}{| c | c | c |}
\hline
\textbf{Language} & \textbf{\%} \\ \hline
English         & 96.31         \\ \hline
French          & 0.86          \\ \hline
German          & 0.86          \\ \hline
Spanish         & 0.62          \\ \hline
Norwegian       & 0.25          \\ \hline
Latin           & 0.25          \\ \hline
Italian         & 0.25          \\ \hline
Welsh           & 0.12          \\ \hline
Turkish         & 0.12          \\ \hline
Portuguese      & 0.12          \\ \hline
Dutch           & 0.12          \\ \hline
Catalan         & 0.12          \\ \hline
\end{tabular}
\label{table:idiomas}
\end{table}

\subsection{Connectivity analysis}

 Several works in the literature have successfully addressed the study of the Surface Web connectivity as a graph~\cite{kleinberg1999web,broder2000graph}. In this section, we will adopt a similar strategy to understand and unveil I2P eepsites relationships and organization patterns. First of all, a brief explanation of graph theory concepts is introduced. Second, specific results will be presented and discussed in detail.
 
Let $G=(V,E)$ be a directed graph comprising a set of nodes $V$ (eepsites) an a set of edges $E$ (connections). Each edge connects a pair of nodes $(u,v)$ through a direct connection from $u$ to $v$. This way, the number of edges from node $u$ to some other nodes: $(u,v_1),(u,v_2), \cdots,(u,v_n)$, is denoted as \textit{out-degree} (eepsite outgoing links). Similarly, the \textit{in-degree} (eepsite incoming links) value of a node $u$ corresponds to the number of edges pointing to node $u$ from some other nodes: $(w_1,u),(w_2,u), \cdots,(w_m,u)$. Based on that, we can define four types of nodes: \textit{source} nodes, with only outgoing links; \textit{sink} nodes, with only incoming links; \textit{connected} nodes, with both incoming and outgoing links; and \textit{isolated} nodes, which are not connected to any other node in the graph. 

\begin{figure}[t]
	\centering
	\includegraphics[width=0.5\textwidth]{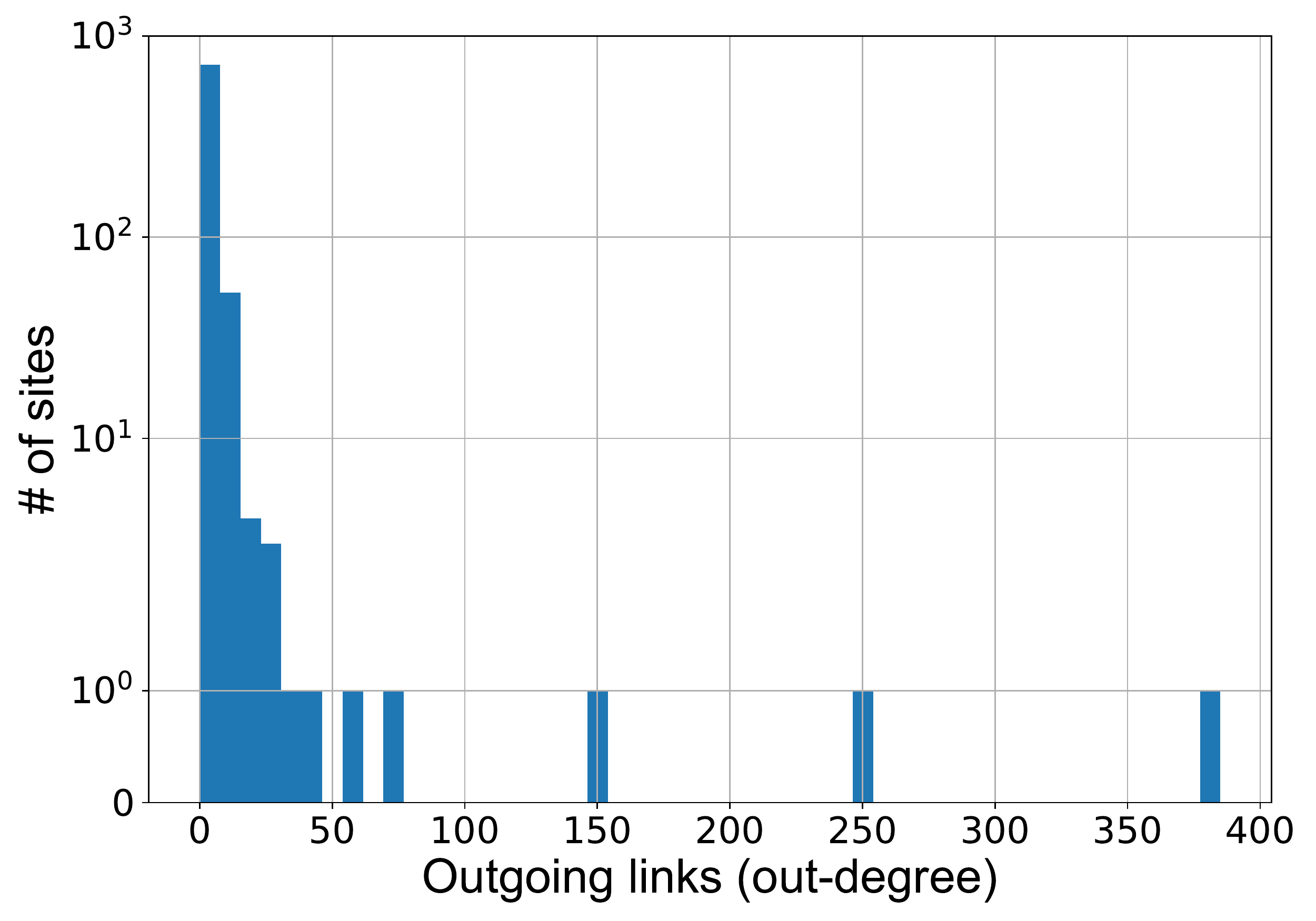}
	\caption{Distribution of eepsite outgoing links (out-degree).}
	\label{fig:histsitesoutgoing}
\end{figure}

\begin{figure}[t]
	\centering
	\includegraphics[width=0.5\textwidth]{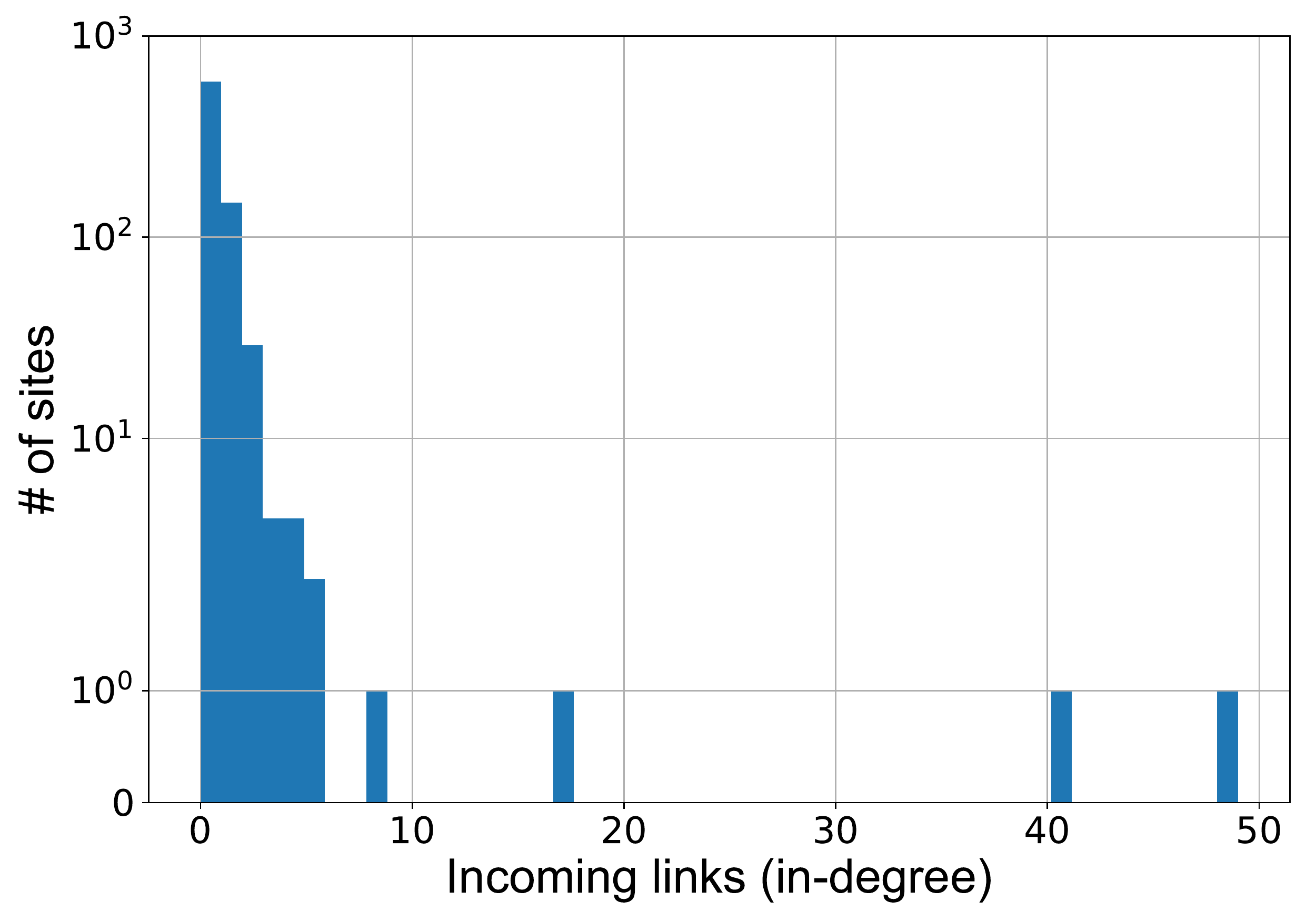}
	\caption{Distribution of eepsite incoming links (in-degree).}
	\label{fig:histsitesincoming}
\end{figure}

The distribution of outgoing and incoming links is shown in Figures~\ref{fig:histsitesoutgoing} and~\ref{fig:histsitesincoming}, respectively. We can observe that most of the crawled nodes have not significant out-degree or in-degree values. In fact, $\sim 10\%$ of the nodes correspond to source nodes, $\sim 12\%$ to sink nodes, and $\sim 12\%$ to connected nodes. Finally, $\sim 66\%$ of nodes are isolated nodes. 

We should remark that there are some eepsites that can be considered as anomalous in terms of out- and in-degree values. Nodes with highest out- and in-degree values are summarized in Tables~\ref{table:top10outgoingparametros} and~\ref{table:top10incomingparametros}, respectively. In particular, the node with the highest value of outgoing links, 385 in total, is the \texttt{identiguy.i2p} eepsite. It contains a dynamically updated list of eepsites but does not cover the whole I2P, i.e. the hidden part. On the other hand, the node with the highest value of incoming links, 58 in total, corresponds with \texttt{forum.i2p}. It is worth noting that most of the biggest in-degree nodes were offline (discarded) probably due to two principal reasons: the number and relationships among the nodes are inconsistent, or many of the eepsites are unattended and not frequently updated.

In addition, all the highest out- and in-degree values corres\-pond to eepsites found from seed source. Based on that, we can confirm the existence and persistence of some relevant I2P eepsites supporting a kind of network backbone among eepsites. However, there is also an important group of eepsites, the isolated ones.


\begin{table*}[t]
\centering
\begin{tabular}{| c | c | c | c |}
\hline
\textbf{\textit{out-degree}} & \textbf{Eepsite} & \textbf{Status} & \textbf{Source}  \\ \hline
385 & \texttt{identiguy.i2p}	& FINISHED      & SEED                       \\ \hline
248 & \texttt{i2pwiki.i2p}	    & FINISHED      & SEED               \\ \hline
152 & \texttt{inr.i2p}	        & FINISHED      & SEED               \\ \hline
70  & \texttt{s6lagaqbn572fvnr7vsxsqwbwxb6m3gr5hu6eetstykdm2opp2ia.b32.i2p}	    & FINISHED  & FLOODFILL                \\ \hline
54  & \texttt{zy37tq6ynucp3ufoyeegswqjaeofmj57cpm5ecd7nbanh2h6f2ja.b32.i2p}	     & FINISHED      & SEED                \\ \hline
45  & \texttt{5ypxuqf2ufqdsf3ejv5xwrgwatjxf2uw7tyxz2av44pka4w3pvza.b32.i2p}	  &   FINISHED  & FLOODFILL                \\ \hline
31  & \texttt{fex6v4zccrovs7dixqbigbbqtrb7ylrmpgphwnwoyjutrg56qmoa.b32.i2p}	  &   FINISHED  & FLOODFILL                \\ \hline
28  & \texttt{i2pforum.i2p}	  &   FINISHED  & DISCOVERED                \\ \hline
26  & \texttt{stats.i2p}	  &   FINISHED  & DISCOVERED                \\ \hline
26  & \texttt{pwgma3snbsgkddxgb54mrxxkt3l4jzchrtp52vxmw7rbkjygylxq.b32.i2p}	  &   FINISHED  & SEED                \\ \hline
19  & \texttt{trac.i2p2.i2p}	  &   FINISHED  & SEED                \\ \hline
18  & \texttt{i2p-projekt.i2p}	  &   FINISHED  & SEED                \\ \hline
17  & \texttt{isxls447iuumsb35pq5r3di6xrxr2igugvshqwhi5hj5gvhwvqba.b32.i2p}	  &   FINISHED  & SEED                \\ \hline
17  & \texttt{4bpcp4fmvyr46vb4kqjvtxlst6puz4r3dld24umooiy5mesxzspa.b32.i2p}	  &   FINISHED  & SEED                \\ \hline
15  & \texttt{uda2rkhskjdb4w7xiftz3btfpl7bhxsy5gwpiiiongte4gulbuza.b32.i2p}	  &   FINISHED  & FLOODFILL                \\ \hline
\end{tabular}
\caption{Eepsites with largest out-degree values.}
\label{table:top10outgoingparametros}
\end{table*}

\begin{table*}[t]
\centering
\begin{tabular}{| c | c | c | c |}
\hline
\textbf{\textit{in-degree}} & \textbf{Eepsite} & \textbf{Status} & \textbf{Source}  \\ \hline
58 & \texttt{forum.i2p}	    & DISCARDED   & SEED                           \\ \hline
55 & \texttt{ugha.i2p}	    & DISCARDED   & SEED                           \\ \hline
52 & \texttt{www.i2p2.i2p}	    & DISCARDED   & SEED                           \\ \hline
49 & \texttt{i2p-projekt.i2p}	    & FINISHED   & SEED                           \\ \hline
47 & \texttt{no.i2p}	    & DISCARDED   & SEED                           \\ \hline
46 & \texttt{inproxy.tino.i2p}	    & DISCARDED   & SEED                           \\ \hline
45 & \texttt{i2host.i2p}	    & DISCARDED   & SEED                           \\ \hline
45 & \texttt{perv.i2p}	    & DISCARDED   & SEED                           \\ \hline
45 & \texttt{tino.i2p}	    & DISCARDED   & SEED                           \\ \hline
41 & \texttt{i2pwiki.i2p}	    & FINISHED   & SEED                           \\ \hline
21 & \texttt{sperrbezirk.i2p}	    & DISCARDED   & SEED                           \\ \hline
17 & \texttt{diftracker.i2p}	    & FINISHED   & SEED                           \\ \hline
13 & \texttt{visibility.i2p}	    & DISCARDED   & SEED                           \\ \hline
12 & \texttt{vstr4d.i2p}	    & DISCARDED   & SEED                           \\ \hline
11 & \texttt{www.imule.i2p}	    & DISCARDED   & SEED                           \\ \hline
11 & \texttt{bote.i2p}	    & DISCARDED   & SEED                           \\ \hline
11 & \texttt{bkillyourtv.i2p}	    & DISCARDED   & SEED                           \\ \hline
\end{tabular}
\caption{Eepsites with highest in-degree values.}
\label{table:top10incomingparametros}
\end{table*}

\begin{figure}[t]
	\centering
	\includegraphics[width=0.5\textwidth]{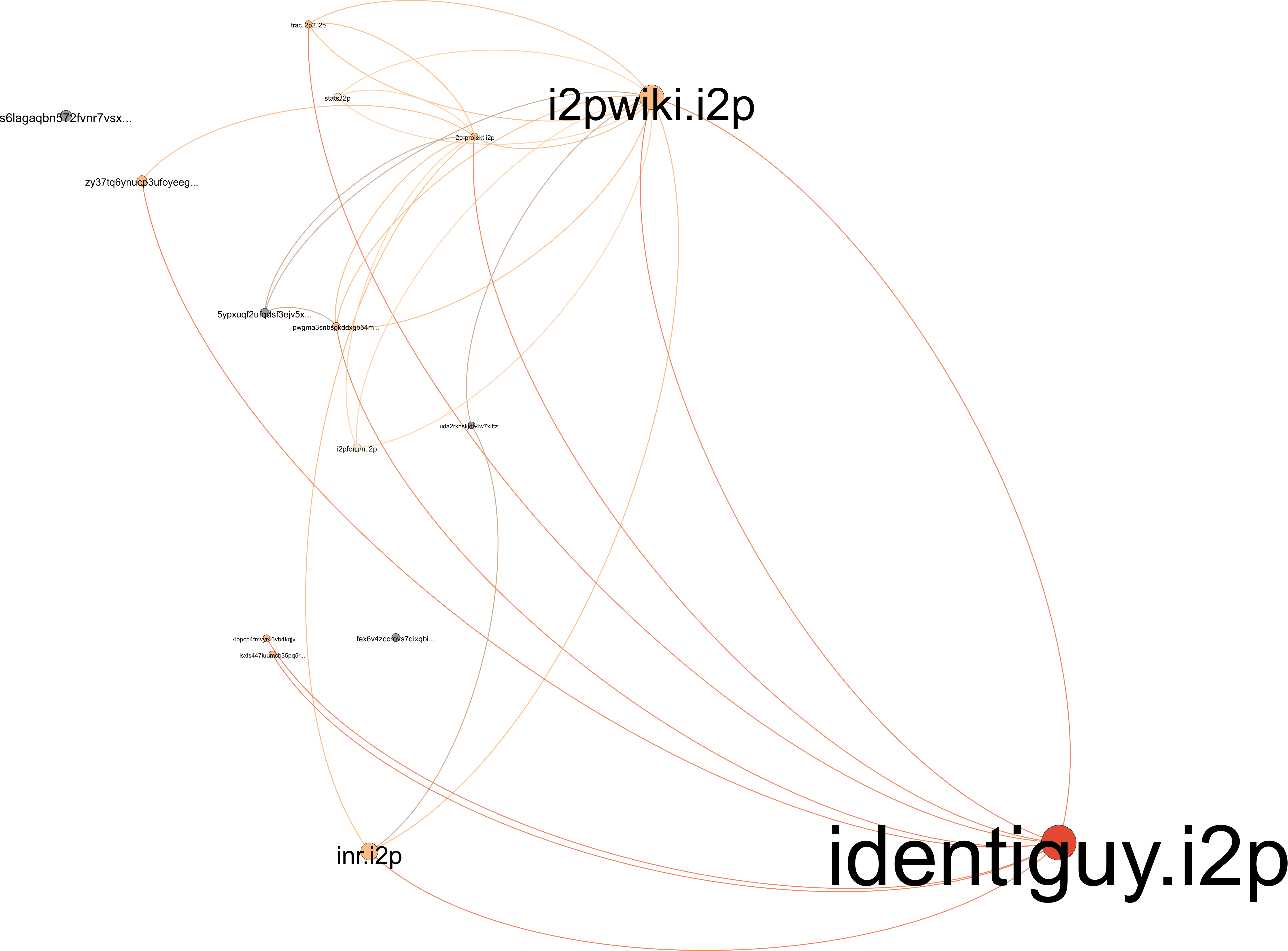}
	\caption{Connectivity graph for nodes with highest out-degree values.}
	\label{fig:grafotop10outgoing}
\end{figure}

\begin{figure}
	\centering
	\includegraphics[width=0.4\textwidth]{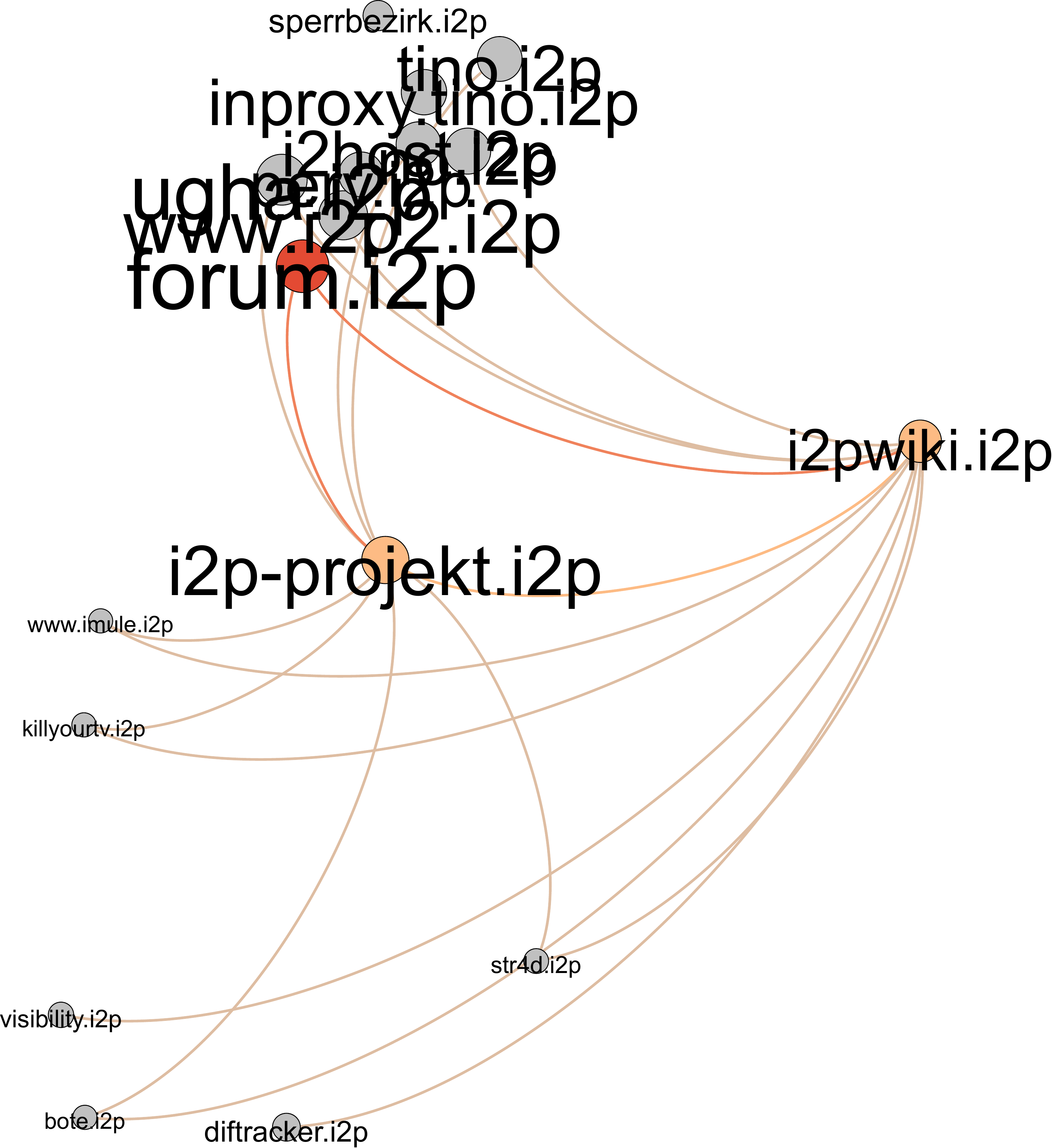}
	\caption{Connectivity graph for nodes with highest in-degree values.}
	\label{fig:grafotop10incoming}
\end{figure}

Figures~\ref{fig:grafotop10outgoing} and~\ref{fig:grafotop10incoming} depict the relationships among top eepsites with the highest out- and in-degree values. In the figures, the size of the labels and nodes is directly proportional to their out- and in-degree values, respectively. As shown, all the nodes are connected. 

\begin{figure*}[t]
	\centering
	\includegraphics[width=1\textwidth]{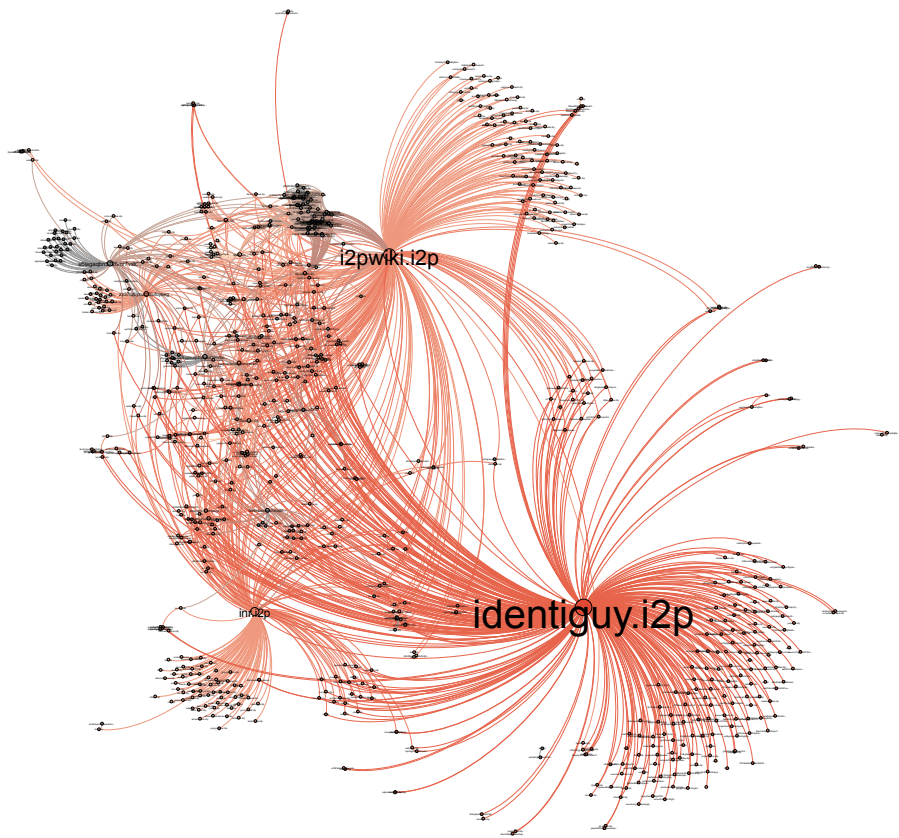}
	\caption{Overall view of the I2P eepsites with at least one incoming or outgoing link.}
	\label{fig:overall_view}
\end{figure*}

An overall view of the connectivity of the I2P network is finally shown in Figure~\ref{fig:overall_view}, where only nodes with at least one incoming or outgoing link are depicted. In this figure, the neighborhood for node \texttt{identiguy.i2p} can be observed, where all the nodes (colored nodes and links) can be accessed \texttt{identiguy.i2p} in three or less hops. It is also remarkable the existence of pairs of nodes separated from the rest.

\section{Conclusions and future work}\label{sec:conclusiones}

Crawling the Surface Web has been widely addressed by the research community with different aims. However, the Deep Web part is still highly unknown despite its big interest not only for researchers, but also for authorities and security forces.
 
In this work, we inspect the I2P darknet by devising new methods and tools based on crawling procedures. The results obtained conclude that I2P is a highly decentralized network, where new services dynamically appear and disappear over time. A small portion of the I2P services are web related services, called eepsites. Regarding content and size aspects, it can be concluded that eepsites are mainly small websites. From the perspective of connectivity and relationships, eepsites are very disconnected from the rest, i.e., they present low in- and out-going connectivity degrees. On the contrary, few of them highlight from the rest in the same terms thus having a kind of eepsite network backbone.  Moreover, more than a half of the total observed eepsites in our experimentation were completely isolated, thus becoming a hidden part of the overall set of eepsites.

Beyond the results obtained, some further work should be done. This way, we plan to extend the experimentation to cover as many I2P nodes as possible. Another relevant focus is to study isolated eepsites and those separated from the rest making groups. It would be also interesting to analyze some other type of services used in I2P. Finally, aimed to get a more in-depth knowledge of the Deep Web, it would be necessary to extend the tool to address other darknet technologies, like Freenet (\url{https://freenetproject.org/}) for a further comparison. 

\section*{Acknowledgement}

This work has been partially supported by Spanish Government-MINECO (Ministerio de Econom\'ia y Competitividad), the ERDF (European Regional Development Fund) through project TIN2017-83494-R (MDSM) and the 5B starting\- research grants program of the University of Granada.

\bibliographystyle{IEEEtran}
\bibliography{refs}

\end{document}